\documentclass[aps,prd,tightenlines,superscriptaddress,nofootinbib,twocolumn]{revtex4-1}

\usepackage[T1]{fontenc}
\usepackage[colorlinks=true]{hyperref}
\usepackage{amsmath}
\usepackage{amssymb}
\usepackage{graphicx}
\usepackage{color}
\usepackage{multirow}
\usepackage{epstopdf}
\usepackage{subfigure}
\usepackage[acronym,shortcuts]{glossaries}
\usepackage{placeins}
\usepackage{cleveref}

\begin{document}

\title{Cost-benefit analysis for commissioning decisions in GEO\,600}

\author{T.~Adams}
\email{thomas.adams@ligo.org}
\affiliation {Laboratoire d'Annecy-le-Vieux de Physique des Particules (LAPP), Universit\'e de Savoie, CNRS/IN2P3, F-74941 Annecy-le-Vieux, France }
\affiliation{School of Physics and Astronomy, Cardiff University, Cardiff, United Kingdom, CF24 3AA}
\author{J.~R.~Leong}
\affiliation{Max-Planck-Institut f\"ur Gravitationsphysik (Albert-Einstein-Institut) und Leibniz Universit\"at Hannover, D-30167 Hannover, Germany}
\author{J.~Slutsky}
\affiliation{ CRESST and Gravitational Astrophysics Laboratory NASA/GSFC, Greenbelt, MD 20771, USA}
\affiliation{Max-Planck-Institut f\"ur Gravitationsphysik (Albert-Einstein-Institut) und Leibniz Universit\"at Hannover, D-30167 Hannover, Germany}
\author{M.~W\k{a}s}
\affiliation {Laboratoire d'Annecy-le-Vieux de Physique des Particules (LAPP), Universit\'e de Savoie, CNRS/IN2P3, F-74941 Annecy-le-Vieux, France }
\affiliation{Max-Planck-Institut f\"ur Gravitationsphysik (Albert-Einstein-Institut) und Leibniz Universit\"at Hannover, D-30167 Hannover, Germany}
\author{C.~Affeldt}
\affiliation{Max-Planck-Institut f\"ur Gravitationsphysik (Albert-Einstein-Institut) und Leibniz Universit\"at Hannover, D-30167 Hannover, Germany}
\author{J.~Degallaix}
\affiliation {Laboratoire des Mat\'eriaux Avanc\'es (LMA), IN2P3/CNRS, Universit\'e de Lyon, F-69622 Villeurbanne, Lyon, France }
\author{K.~L.~Dooley}
\affiliation{Max-Planck-Institut f\"ur Gravitationsphysik (Albert-Einstein-Institut) und Leibniz Universit\"at Hannover, D-30167 Hannover, Germany}
\author{H.~Grote}
\affiliation{Max-Planck-Institut f\"ur Gravitationsphysik (Albert-Einstein-Institut) und Leibniz Universit\"at Hannover, D-30167 Hannover, Germany}
\author{S.~Hild}
\affiliation {SUPA, University of Glasgow, Glasgow, G12 8QQ, United Kingdom }
\author{H.~L\"uck}
\affiliation{Max-Planck-Institut f\"ur Gravitationsphysik (Albert-Einstein-Institut) und Leibniz Universit\"at Hannover, D-30167 Hannover, Germany}
\author{D.~M.~Macleod}
\affiliation {Louisiana State University, Baton Rouge, LA 70803, USA }
\author{L.~K.~Nuttall}
\affiliation {University of Wisconsin--Milwaukee, Milwaukee, WI 53201, USA }
\author{M.~Prijatelj}
\affiliation {European Gravitational Observatory (EGO), I-56021 Cascina, Pisa, Italy }
\author{E.~Schreiber}
\affiliation{Max-Planck-Institut f\"ur Gravitationsphysik (Albert-Einstein-Institut) und Leibniz Universit\"at Hannover, D-30167 Hannover, Germany}
\author{B.~Sorazu}
\affiliation {SUPA, University of Glasgow, Glasgow, G12 8QQ, United Kingdom }
\author{K.~A.~Strain}
\affiliation {SUPA, University of Glasgow, Glasgow, G12 8QQ, United Kingdom }
\author{P.~J.~Sutton}
\affiliation{School of Physics and Astronomy, Cardiff University, Cardiff, United Kingdom, CF24 3AA}
\author{H.~Vahlbruch}
\affiliation{Max-Planck-Institut f\"ur Gravitationsphysik (Albert-Einstein-Institut) und Leibniz Universit\"at Hannover, D-30167 Hannover, Germany}
\author{H.~Wittel}
\affiliation{Max-Planck-Institut f\"ur Gravitationsphysik (Albert-Einstein-Institut) und Leibniz Universit\"at Hannover, D-30167 Hannover, Germany}
\author{K.~Danzmann}
\affiliation{Max-Planck-Institut f\"ur Gravitationsphysik (Albert-Einstein-Institut) und Leibniz Universit\"at Hannover, D-30167 Hannover, Germany}

\begin{abstract}
Gravitational wave interferometers are complex instruments, requiring years of commissioning to achieve the required sensitivities for the detection of gravitational waves, of order $\lesssim10^{-21}$ in dimensionless detector strain, in the tens of Hz to several kHz frequency band.
Investigations carried out by the GEO\,600 detector characterisation group have shown that detector characterisation techniques are useful when planning for commissioning work.
At the time of writing, GEO\,600 is the only large scale laser interferometer currently in operation running with a high duty factor, $\sim70\%$, limited chiefly by the time spent commissioning the detector.
The number of observable gravitational wave sources scales as the product of the volume of space to which the detector is sensitive and the observation time, so the goal of commissioning is to improve the detector sensitivity with the least possible detector down-time.
We demonstrate a method for increasing the number of sources observable by such a detector, by assessing the severity of non-astrophysical noise contaminations to efficiently guide commissioning.
This method will be particularly useful in the early stages and during the initial science runs of the aLIGO and adVirgo detectors, as they are brought up to design performance.
\end{abstract}

\maketitle

\section{Introduction}
\label{sec:intro}
GEO\,600 \cite{Grote:2010tg} is one member of a global network of interferometric gravitational wave (GW) detectors aiming to make the first direct observation of GWs along with the LIGO \cite{Abbott:2009li}, Virgo \cite{Accadia:2012vi}, and KAGRA \cite{Somiya:2012en} detectors.
Current state of the art laser interferometric GW detectors are instruments designed to detect GWs between $\sim10$\,Hz and a few kHz.
GEO\,600 is a German-British detector located near Hannover, Germany.
The optical set up of GEO\,600 is a dual recycled optical folded arm Michelson interferometer with $600$\,m long beam lines.
At the time of writing GEO\,600 is the only operating large scale interferometer with high duty cycle.
Since 2010 a series of upgrades, referred to as the GEO-HF scheme, has been under way to improve the sensitivity to GWs at frequencies above a few hundred Hz \cite{Willke:2006ge}.
The upgrades constituting the GEO-HF program include increasing the laser power and the implementation of squeezed light \cite{Vahlbruck:2010sq}.
During this period of upgrades GEO\,600 aims to maintain an average duty cycle of $\sim70\%$, while the remaining $\sim30\%$ of time is mostly spent on commissioning activities.

The frequency-dependent sensitivity of a GW detector is limited by a stationary noise floor overlaid with non-stationary transient noise artefacts.
These components limit the sensitivity to GWs signals in two ways: a high noise floor masks GWs in the data (reducing the signal-to-noise ratio (SNR) of signals), while transients can be confused with GWs.
These transients produce a substantial background of candidate events that limit the confidence of GW identification; therefore, we require higher amplitude GWs for confident identification.
The goal of detector characterisation is to identify causes of non-stationary noise and the components of the noise floor, in order to reduce both.

GW detectors are complicated instruments made up of dozens of subsystems and control loops.
Characterising the detector behaviour works to aid both analysis of collected data and commissioning of the detector.
The GEO\,600 detector characterisation (GEODC) group has focused mainly on aiding commissioning work, through clear and frequent communication with the commissioning team, which is powerfully aided by having GEODC members on-site at the detector.
The GEODC group investigates noise artefacts in order to understand and eliminate the physical causes.
Information is fed back to the on-site commissioning team who carry out investigations, fix hardware issues, and thereby improve the instrument sensitivity.
The efforts of the GEODC group have played a major role in investigations of instrumental artefacts and the removal of their sources.

We aim to increase the number of potential GW signals that are observable in our data.
This number is proportional to both the volume of space containing canonical signals we are sensitive to and the amount of time for which we are observing.
By commissioning the detector we aim to increase the sensitive range, and therefore increase the volume of space we are observing.
However, commissioning interrupts the data taking which reduces the time spent observing, this time is referred to as ``down-time''.
Every investigation costs observing time, so it is important to select them in order to maximize the probability of GW detection.

In this paper we report on a selection of transient noise classes that the GEODC group identified between February 2011 and March 2013.
The layout of this paper is as follows.
Section \ref{sec:tools} defines the diagnosis tools used for the detector characterisation investigations, and
Section \ref{sec:procedure} describes the procedure for performing these investigations.
Section \ref{sec:geo_layout} briefly describes the key elements of GEO\,600 that are referred to in the investigations described in this paper.
Section \ref{sec:investigations} explains each noise phenomenon investigated.
For each investigation, veto and commissioning solutions are analysed, when available, and we use our diagnostic tools to quantify this performance.
Section \ref{sec:summary} summarises the investigation results, and how these methods should be useful for future commissioning efforts.

\section{Diagnostic tools}
\label{sec:tools}
In order to make informed decisions about which commissioning interruptions can increase the probability of GW detection, it is important to characterise and assess the severity of noise sources that are observed in the output data stream of the detector.
In this paper we use a variety of the standard characterisation software tools for this purpose which we describe in this section, however, we have developed here a method for utilising them to make astrophysically motivated commissioning decisions during scientific observing runs of a GW detector.

We compare initial periods, both with and without the use of any available vetoes to reduce the impact of the noise in an astrophysical search, with predicted and/or real post-commissioning periods.
This \emph{before} verses \emph{after} approach is how we assess the severity of a noise source.
This informs us of the impact of any planned or performed commissioning interruption.
The ``predicted'' scenario, in which the effect of a potential element of commissioning is evaluated in advance, enables decisions to be made before performing any commissioning.
We give results also for the actual impacts of commissioning improvements on GEO\,600 as a comparison and reflection of whether the right choice was made where possible.
For this paper, however, all of the predicted scenarios were created after the fact to demonstrate this procedure using our knowledge at the time of the noise source to make our prediction.
The use of vetoes in the initial periods will demonstrate how well we can do at removing the noise without commissioning, and in some cases offers an alternative solution which allows us to relax the need of immediate commissioning.

\begin{description}
\item[Strain noise spectral density] is the classic characterisation of stationary noise levels in the detector as a function of frequency.
When the noise is non-stationary, the noise spectral density represents the stationary components of the noise plus a particular average of the transients over the time interval of the measurement.
The spectrum gives the base noise level to which any transient (signal or noise) must be compared in order to rank its significance.
It is sensitive to detector configuration changes or egregious transients which are either very loud or occur very often.
See figure~\ref{fig:geo_spectra} for examples.

\item[Omegamaps] give a multiple quality-factor, time-frequency representation of the detector data (for example figure \ref{fig:bdo_omegamap}).
However, unlike a spectrogram Omegamaps can simultaneously show information about both short-duration, broadband; and long-duration, narrow-band structures.
The map is normalised by the strain noise spectral density to highlight transient deviation from it.
At GEO\,600 we found this visualisation method to be almost as good as the human ear at picking out transient events and also added a quantitative element which the ears lacked.
This tool is used to give a hint at which directions to drive the commissioning investigations and development of vetoes.

\item[Event trigger generators] produce lists of transient noise events for the detector strain data channel and a large number of auxiliary channels.
This is done by thresholding time-frequency maps similar to the Omegamaps.
These events are referred to as ``triggers''.
The triggers for the detector strain data channel are used as one of the inputs to GW searches which attempt to separate out those which come from  signals and those which come from noise.
At GEO\,600 both the hierarchical algorithm for curves and ridges (HACR) \cite{Heng:2004tv} and the Omega-pipeline \cite{Chatterji:2005th,Parameswaran:2007ve,Rollins:2011tc} are used to produce triggers.
Parameters of the transient are reported, such as the central frequency, bandwidth, peak time, duration, and SNR.
These trigger generators, as well as the related Coherent WaveBurst algorithm used in \cref{sec:dither_glitches}, perform similar analyses tasks using different algorithms, and have been used as convenient by the GEODC team primarily based on our familiarity with them, and on the availability of support from their authors.
In principle, a single trigger generator could be used, and in practice we relied mostly on the Omega-pipeline for the diagnostic tools listed below.

\item[Cumulative glitch rate histograms] illustrate the cumulative SNR rate distribution of triggers (for example figure \ref{fig:bdo_trig_rate}).
When used on triggers which are known to be ``noise'' with respect to a particular search, these histograms can be used along with the length of an observation block to calculate the false-alarm probability of a candidate signal trigger with a given SNR \cite{Was:2014fo}.
They thus serve as a characterisation of the non-stationary noise in the detector and are a necessary intermediate data product of every search algorithm.

\item[Fixed false-alarm probability range] combines information from the stationary and non-stationary noise characterisations, the strain noise spectral density and the cumulative glitch rate histogram respectively, into a single number which is a measure of the farthest distance at which a gravitational wave burst source of a given energy could be detected in a detector.
See \cite{Was:2014fo} for a full description.

In general the formulation of the range at a given frequency $f$ is~\cite{Sutton:2013uf}
\begin{equation}
    R(f) = \left(\frac{G E_{GW}}{2 \pi^{2} c^3}\right)^{1/2}  \frac{1}{\rho \sqrt{S(f)} f} \, ,
    \label{eq:freq_dep_range}
\end{equation}
where $E_{GW}$ is the energy released in GWs; $\rho$ is an SNR detection threshold; and $S(f)$ is the strain noise spectral density.

To find the range for a particular frequency band, for example the band in which we expect GWs for a particular source, we define an integrated range.
For a frequency band $f_{1}$ to $f_{2}$ we define the integrated range \cite{Was:2014fo} as
\begin{equation}
    \left< R \right> \equiv \left[ \frac{1}{f_{2}-f_{1}} \int^{f_{2}}_{f_{1}} R(f_{0})^{3}df_{0} \right]^{1/3} \, .
    \label{eq:integrated_range}
\end{equation}

The fixed false-alarm probability range is calculated using the definition in equation \ref{eq:integrated_range} with $\rho$ defined by the SNR corresponding to a pre-determined false-alarm probability level, $\alpha$, in the cumulative glitch rate histogram.
Here all background triggers that overlap the frequency range $f_{1}$ to $f_{2}$ are used to calculate the histogram.
Using an SNR defined in this manner assures that we always define our detection threshold to correspond with the given probability, $\alpha$, of the detection being a false positive.

In practice we use parameters corresponding to two putative searches for gravitational wave bursts: those emitted from a nearby gamma-ray burst (GRB) or those emitted from a galactic supernova (SN).
For the GRB case we assume an emission of $E_{GW}=10^{-2}M_\odot c^2$ in the frequency band of $100\,\mathrm{Hz}$--$500\,\mathrm{Hz}$, for the SN case we assume an emission of $E_{GW}=10^{-8}M_\odot c^2$ in the frequency band of $500\,\mathrm{Hz}$--$4\,\mathrm{kHz}$.
In both cases we use an SNR threshold corresponding to a false alarm probability of $\sim10^{-3}$ in a few seconds long search window around the neutrino or gamma-ray observation of the SN or GRB, with a corresponding false alarm rate of $5.6\times10^{-4}\,\mathrm{Hz}$.
These two ranges give us the basic astrophysical figure-of-merits for the low and high frequency sensitivity of GEO\,600.

\item[Spacetime observation rate] is an extension of the fixed false-alarm probability range which measures the rate at which we are accumulating observed spacetime volume for a particular source.
This is derived from the quantity of most interest to our astrophysical searches: the expected number of observed sources, $N_{\mathrm{GW}}$, during a period of observation of length $t$.
For a uniform distribution of sources, $B$, as in \cite{Was:2014fo} we have
\begin{equation}
    N_{\mathrm{GW}} = \frac{4\pi}{3}\left< R \right>^{3} B t \, .
    \label{eq:nos}
\end{equation}
Expanding the observation time, $t$, as the product of the total time, $T$, and an operation duty cycle, $D$, we can then calculate the spacetime observation rate as
\begin{equation}
    \dot{S} \equiv \frac{N_{\mathrm{GW}}}{B T} = \frac{4\pi}{3}\left< R \right>^{3} D \, .
    \label{eq:star}
\end{equation}
This is the basic figure-of-merit we will use to compare two configurations of data production in a GW detector.
It combines information about the sensitivity of the detector through the fixed false-alarm probability range, $\left< R \right>$, with the duty cycle of data production, $D$, hindered either by regular detector down-times or time-domain vetoes.

\begin{figure}
    \centering
    \includegraphics[width=0.45\textwidth]{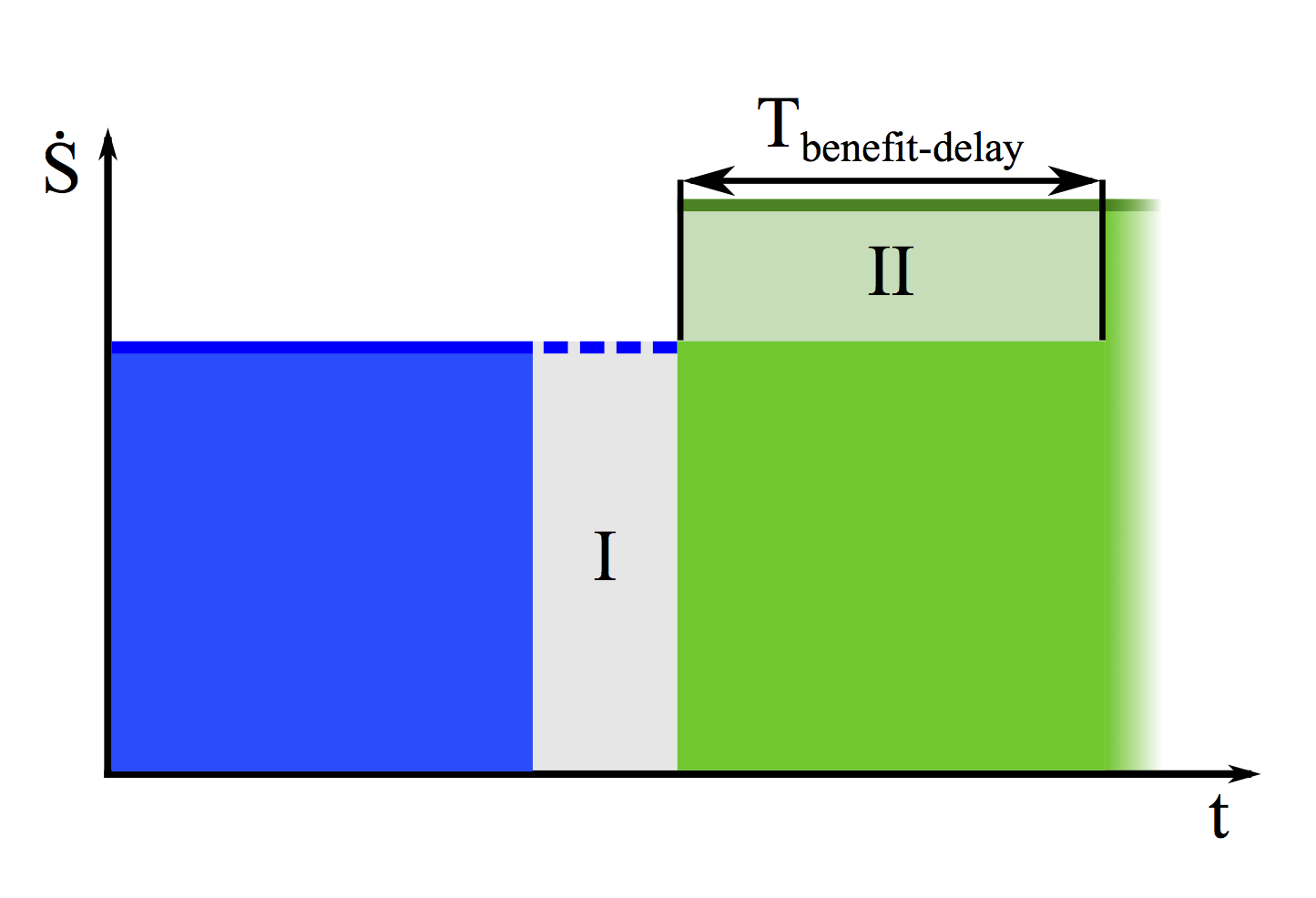}
    \caption{\label{fig:fom}Cartoon showing the spacetime observation rate ($\dot{S}$) against time for a noise investigation.
             As defined in equation \ref{eq:star}, areas under this curve are proportional to the expected number of observed sources.
             The blue area denotes a period of detector performance while the noise was present before the commissioning of an improvement occurs.
             In green we show a period after the noise was removed.
             The grey area under the dashed spacetime observation rate denoted by ``$\mathrm{I}$'' represents the potential observed sources which were lost due to the break in operation caused by commissioning.
             The benefit-delay time $T_{\textrm{benefit-delay}}$, is then defined as the time required, while running the detector at the improved spacetime observation rate after commissioning, to make up for the commissioning loss.
             As the area denoted by ``$\mathrm{II}$'' represents the improvement in spacetime observation rate, $T_{\textrm{benefit-delay}}$ is the time which makes $\mathrm{Area}(\mathrm{II}) = \mathrm{Area}(\mathrm{I})$.}
\end{figure}

\item[Vetoes] consist in removing from the analysis triggers that are in a contaminated time-frequency regions.
These contaminated time-frequency regions are derived from investigations and auxiliary sensors in the detector.
We use two types of vetoes: time-domain vetoes and frequency-domain vetoes.
Vetoes remove glitches and hence reduce the $\rho$ threshold in equation \ref{eq:freq_dep_range}, however this comes at a cost that depend on the veto type.

Time-domain vetoes remove triggers with central times within specific slices of time.
Hence they reduce the effective duty cycle $D$.
Frequency-domain vetoes remove triggers with central frequency within specific frequency bands.
Hence they render the detector blind at some frequencies, the range $R(f)$ in equation \ref{eq:integrated_range} is set to zero in the vetoed frequency bands.

The spacetime observation rate includes both the benefits and the cost of using vetoes.

\item[Benefit-delay time and benefit-delay ratio] are figures-of-merit which help us determine how long we need to operate a detector at an improved sensitivity to make up the deficit number of observable sources during the period when the detector is off-line for commissioning.

The benefit-delay time, $T_{\textrm{benefit-delay}}$ is depicted in figure \ref{fig:fom}.
Here we see that
\begin{equation}
    T_{\textrm{benefit-delay}} = \frac{\dot{S_{\textrm{b}}} t_{\textrm{comm}}}{\dot{S_{\textrm{a}}} - \dot{S_{\textrm{b}}}}
    \label{eq:benefit-delay}
\end{equation}
where $t_{\textrm{comm}}$ is the estimated down-time due to commissioning and $\dot{S_{\textrm{b}}}$ and $\dot{S_{\textrm{a}}}$ are the spacetime observation rates before and after the commissioning period respectively.
It should be noted that this definition assumes that the spacetime observation rate is constant enough before the commissioning interruption that the deficit number of observable sources caused by commissioning is proportional to $\dot{S_{\textrm{b}}} t_{\textrm{comm}}$.

It is useful to express the benefit-delay time figure-of-merit as the ratio
\begin{equation}
    \Theta_{\textrm{benefit-delay}} = \frac{T_{\textrm{benefit-delay}}}{t_{\textrm{comm}}} \, ,
    \label{eq:benefit-delay_ratio}
\end{equation}
so that the benefit-delay time can be easily calculated from any commissioning down-time.

These benefit-delay time figures-of-merit are the quantities which we will use to drive commissioning decisions.
If the benefit-delay time is shorter than the remaining time for an observing period after a commissioning break, then the commissioning will increase observable spacetime during the observing period.
If this is not the case, the loss in observable spacetime due to commissioning will not be re-accumulated during the observation period.
For this reason, it is important to estimate $t_{\textrm{comm}}$ conservatively, taking the most pessimistic reasonable estimate into account.
\end{description}

\section{Procedure for Investigations}
\label{sec:procedure}

The broad layout of each investigation in section \ref{sec:investigations} is the same, although details vary in each due to the specifics of the situations.
This section describes the common approach, and how to use the tools described in section \ref{sec:tools} to plan detector commissioning.
The investigations presented in this article were analysed retrospectively, except for the final case of the $3.5\,\mathrm{Hz}$ dither-squeezing glitches, in section \ref{sec:dither_glitches}, in which a prototypical analysis was carried out at the time to make just such a commissioning decision.
We nonetheless outline a procedure for solving noise problems, as they arise, in a manner which is most efficient with respect to the astrophysical productivity of a GW detector.

Two standard techniques are available to mitigate noise contamination of the data.
The first approach is to remove the source of the noise through commissioning the detector.
The second approach is to excise the data affected in the contaminated time-frequency band from analysis, referred to as vetoing.
Vetoes can be applied after the fact, and do not incur the substantial additional periods of down-time from commissioning.
However, effective vetoes are not always available if the phenomenon is severe enough to be endemic or subtle enough to evade automatic veto generation.
Commissioning removes the noise source, meaning that the only lost data is from a one-time commissioning effort.
Unfortunately, the down-time required for investigations and commissioning can be extensive and difficult to predict, and successful noise reduction is not assured.

Initial identification of a problematic noise source is generally through the effect on the spacetime observation rate.
This either results from the effects on the strain noise spectral density, the cumulative glitch rate histogram, or both, depending on the stationarity of the noise source.
These scenarios are distinguished by plotting the strain noise spectral density and the cumulative glitch rate histogram as in figure \ref{fig:bdo_range_inputs}.
Non-stationarity is further exposed by Omegamaps as in figure \ref{fig:bdo_omegamap}, which show the transient nature of the noise, providing more information to assist the determination of the origin of the noise, and also contributing to the possible development of any veto.

After identification of the cause of the noise, investigations on the data can proceed, initially without interfering with detector operation.
During this time planning for experimental tests and possible commissioning solutions are made.
These investigations build an understanding of the noise extent so as to predict the improvement in spacetime observation rate without the noise, and when possible also to produce a veto.
Additionally, short investigations with the instrument are often necessary to understand the noise well enough to generate these elements.

We can then apply the benefit-delay time formalism, defined in section \ref{sec:tools} by equation \ref{eq:benefit-delay} and equation \ref{eq:benefit-delay_ratio}.
Here we use the predicted spacetime observation rate for the ``after commissioning'' scenario to quantify the expected benefit from commissioning the noise source.
If a veto can be constructed, we then calculate the spacetime observation rate with the veto applied and substitute the ``before commissioning'' scenario in the benefit-delay time analysis with this vetoed scenario.
The comparison of the spacetime observation rates from these analyses results in a measure of how effective any available vetoes can be in mitigating the need for commissioning.

For the rest of this article we work through the general method by describing important noise contaminations, and the associated commissioning campaigns in GEO\,600.
For two of the cases in this article, we have easily comparable before- and after-commissioning scenarios where only one problem was addressed during the commissioning down-times.
In situations like this, it is possible to validate the benefit-delay time analysis, to reveal if the commissioning decision was appropriate.
For the final case, we discuss the use and investigate the correctness of a prototype version of this process that was applied in order to decide against aggressive commissioning.

\section{GEO\,600 overview}
\label{sec:geo_layout}

\begin{figure}
    \centering
    \includegraphics[width=0.5\textwidth]{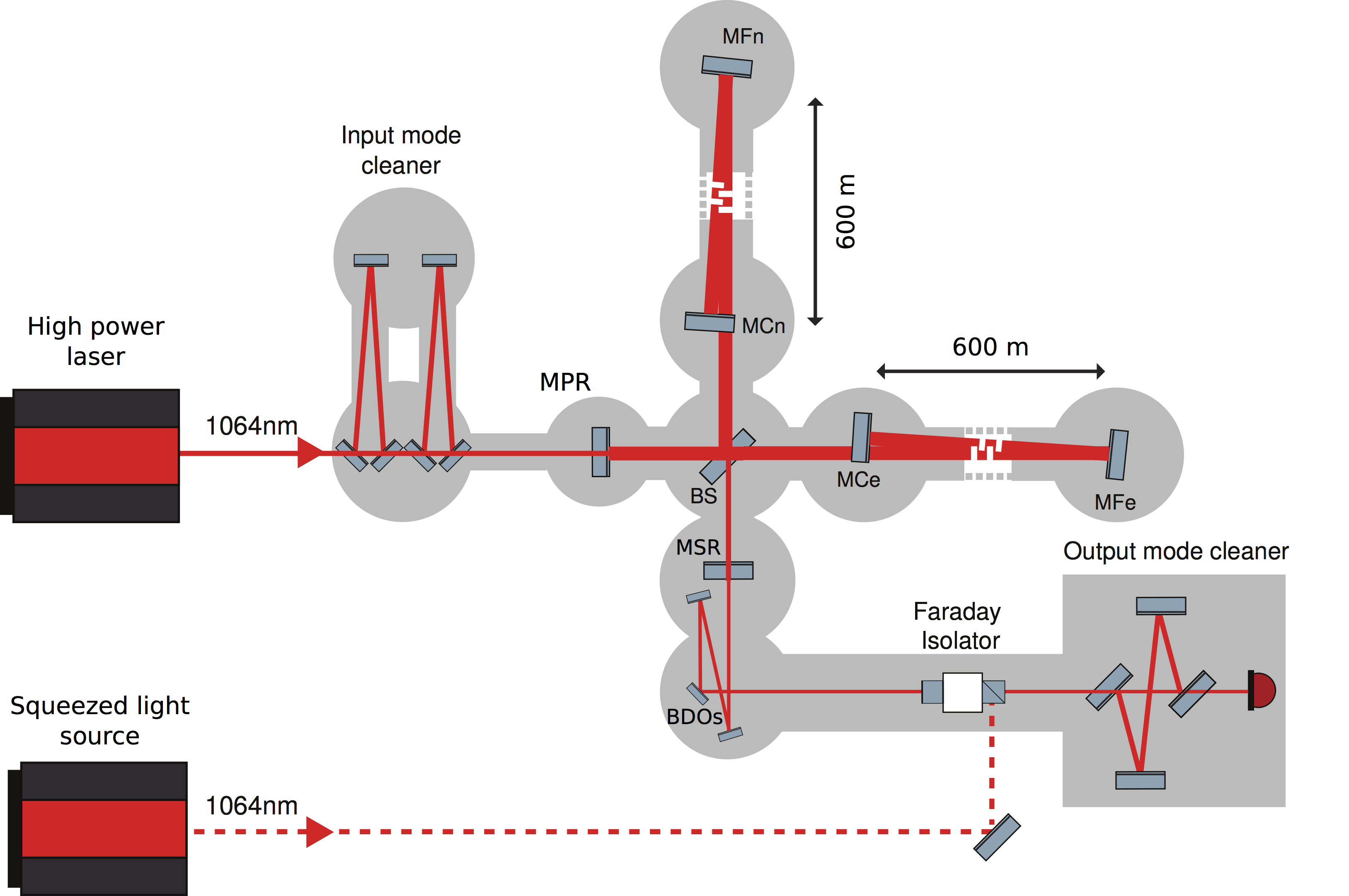}
    \caption{\label{fig:geo_layout}Simple layout of GEO\,600, see section \ref{sec:geo_layout} for details.
             Included are the BDOs, OMC, and squeezer which are relevant for the investigations in section \ref{sec:investigations}.}
\end{figure}

Before going into the investigations in section \ref{sec:investigations}, it is important to give a brief description of some of the relevant elements of GEO\,600.
 A simple schematic of the optical layout of GEO\,600 is illustrated in figure \ref{fig:geo_layout}.
Laser light at a wavelength of $1064\,\mathrm{nm}$ is fed into the vacuum system from an amplitude and frequency stabilised laser.
It first passes through two suspended triangular optical cavities that remove higher order spatial modes of the laser light and reduce amplitude noise.
These cavities are called the input mode cleaners.

The interferometer itself is composed of seven large suspended main optics, which form a power and signal recycled Michelson interferometer with $1200$\,m long folded arms.

At the output of the interferometer, three beam directing optics (the BDO mirrors) are used to direct the light into a Faraday isolator and through an output mode cleaner (OMC).
The purpose of the BDOs is twofold: they form a telescope which reduces the large beam waist of the interferometer to match the small waist of the output optics while also serving as alignment actuators.

\begin{figure*}
    \centering
    \includegraphics[width=0.90\textwidth]{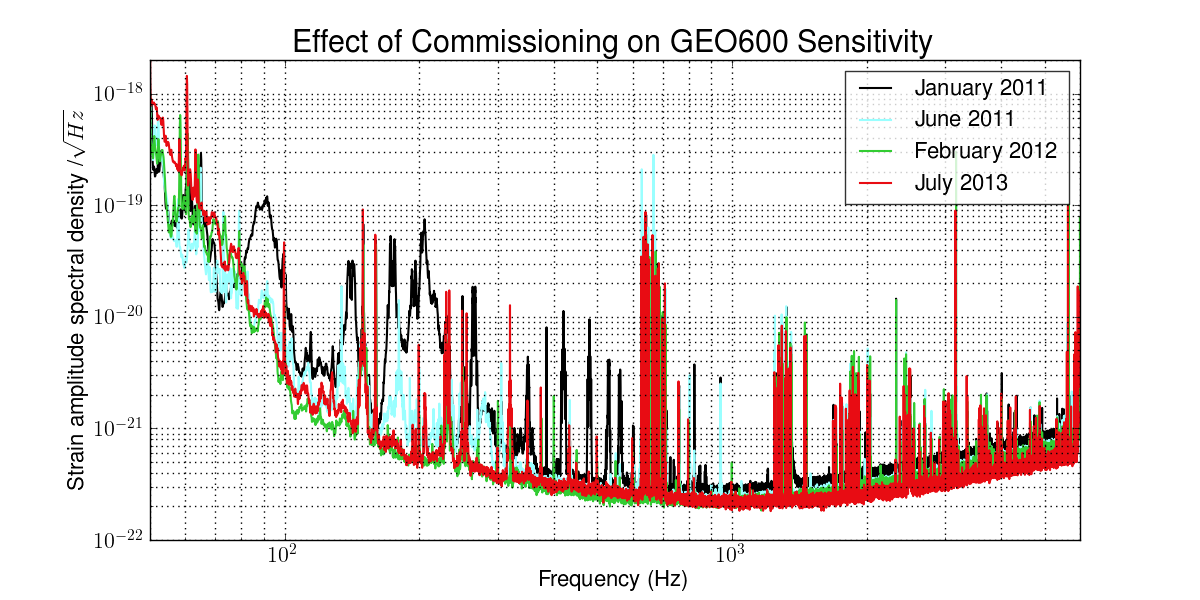}
    \caption{\label{fig:geo_spectra}Strain noise spectral density versus frequency for GEO\,600 for different epochs referred to in section \ref{sec:investigations}.
            In black is depicted the state before the characterisation and commissioning campaign of this paper; in cyan after introducing additional seismic isolation stages to the BDOs; in green after additional isolation of the OMC; and in red after removal of other transient noises and narrow line features, as well as improvement of squeezer performance.}
\end{figure*}

The OMC rejects radio frequency side-bands of the laser light that are used for interferometer control and also rejects higher order spatial modes coming from defects of the main optics.
The power in this \emph{cleaned} light exiting the OMC is detected by the photo-diode shown.
This signal is used to measure changes in the relative arm length inside the interferometer, and is referred to as the strain data channel.

The Faraday isolator is required in the procedure used for the injection of a squeezed vacuum state into the output port of the interferometer.
This squeezed light decreases the shot noise at the photo-diode, the dominant source of noise in the relative arm length sensing at high frequencies (see section \ref{sec:squeezer} for details).
The sensitivity of GEO\,600 can be seen in figure \ref{fig:geo_spectra} where the amplitude spectral density of detector strain is plotted for different epochs that span the work reported in this paper.
The amplitude spectral density is just one way of quantifying the astrophysical sensitivity of a GW detector.

\section{Investigations}
\label{sec:investigations}

This section details three investigations into noise sources that occurred at GEO\,600.
This is not an exhaustive summary of the work done by either the detector characterisation or commissioning teams.
The data used in this paper is not representative of the general sensitivity of GEO\,600, rather these investigations were chosen to display a variety of noise sources requiring different balances of vetoes and commissioning.
The ``before'' and ``after'' commissioning times for each investigation were chosen to highlight the problems clearly, so ideally only the effects of the noise sources being investigated are present.
However, there are many noise sources whose effects on the strain data channel stream show up and disappear unpredictably so this will not always be the case.

A key element for why these investigations were chosen is their ability to demonstrate how a close relationship between a characterisation team and a commissioning team can improve the scientific output of a GW detector.
Important insights are obtained by retroactively applying the robust analysis outlined in section \ref{sec:procedure} to key noise investigations, showing quantitatively how detector characterisation can and should contribute significantly to the direction of the detector commissioning effort; resulting in an overall increase in the number of observable sources compared to unguided commissioning.

The remainder of this sections details our investigations into different noise sources observed at GEO\,600.
Each investigation discussion describes the noise phenomena, veto and commissioning results, compares their performance, and discusses the best solution and other lessons learned.

\subsection{OMC alignment issues}
\label{sec:omc_alignment}

\begin{figure}
    \centering
    \includegraphics[width=0.45\textwidth]{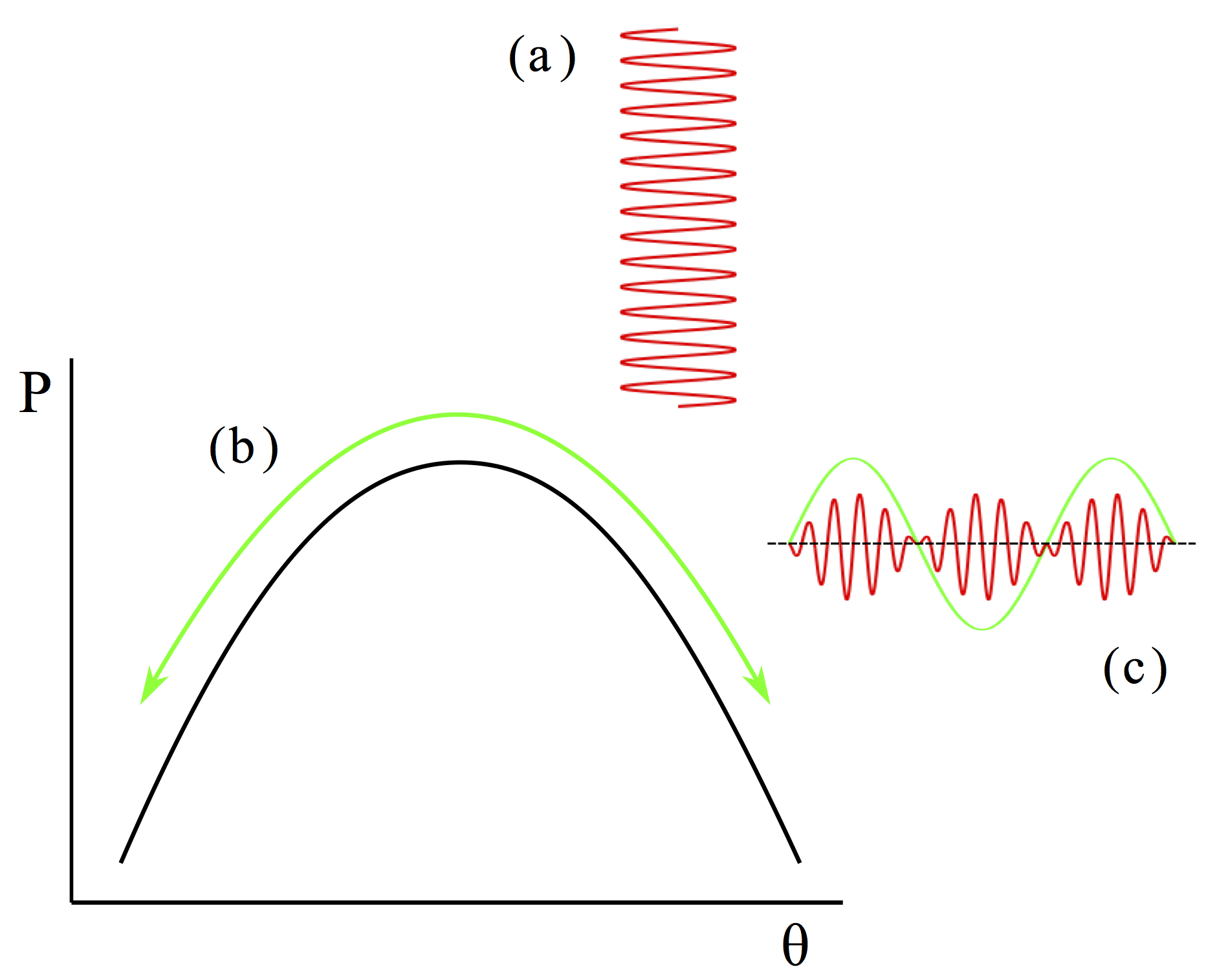}
    \caption{\label{fig:modulation}The transmission power, $P$, of the beam exiting the OMC as a function of an input beam alignment degree of freedom parameter $\theta$ is shown as the black curve.
             A high frequency fluctuation in $\theta$ shown in red (a) is modulated by a low frequency variation shown in green (b).
             This produces a modulated high frequency fluctuation in the transmission (c) shown in red with the low frequency window shown in green.}
\end{figure}
The three optics that match the main interferometer to the OMC are referred to as the BDOs.
Originally the BDO suspensions were not designed to be compatible with an OMC for DC readout, so they were hung as single pendulum suspensions from blade springs, designed to provide isolation in the vertical direction.
These springs are mounted on a rigid support within the vacuum chamber and are effectively connected to the ground at the frequencies of interest.
Due to this single stage design, ground motion easily excites resonances of this system, coupling into mirror motion and ultimately misalignment of the beam impinging on the OMC.

The transverse motion of the suspension point of the optics hung on pendula is highly suppressed at frequencies above the resonance frequencies of the pendula.
At frequencies well below the resonance frequencies there is no such suppression, and there is amplification near the resonances.
At GEO\,600, for the main interferometer the optic's suspensions have resonance frequencies spanning $0.5\,\mathrm{Hz}$--$2\,\mathrm{Hz}$.
At these frequencies there are numerous uncontrolled interferometer degrees of freedom.
This leads to slow drifts of, for example, the position and angle of the output beam of the interferometer.
In turn, this motion misaligns the output beam onto the OMC.

Any additional high frequency alignment fluctuations are modulated by these low frequency variations, producing a noise source that varies in amplitude as the alignment of the OMC changes.
This mechanism is shown in figure \ref{fig:modulation}.
When the OMC is misaligned, a fluctuation entering the OMC will couple into the amplitude transmission of the OMC, and thus the observed detector strain signal, at the frequency of the fluctuation.
Conversely, when the OMC is well aligned, the entering fluctuation will couple more weakly and only to the higher order harmonics of the frequency of fluctuation.

Of particular concern were the violin modes of the steel wires suspending the BDOs which spanned frequencies from $140\,\mathrm{Hz}$--$205\,\mathrm{Hz}$.
Resonances of the support structure from which BDOs are suspended spanned a range of frequencies from $80\,\mathrm{Hz}$--$600\,\mathrm{Hz}$.
Through the mechanism described above, these high frequency fluctuations that show up in the alignment of the OMC were modulated by the large low frequency variation due to the motion of the main optics.

\begin{figure*}
    \centering
        \subfigure[ Strain noise spectral density]{\includegraphics[width=0.45\textwidth]{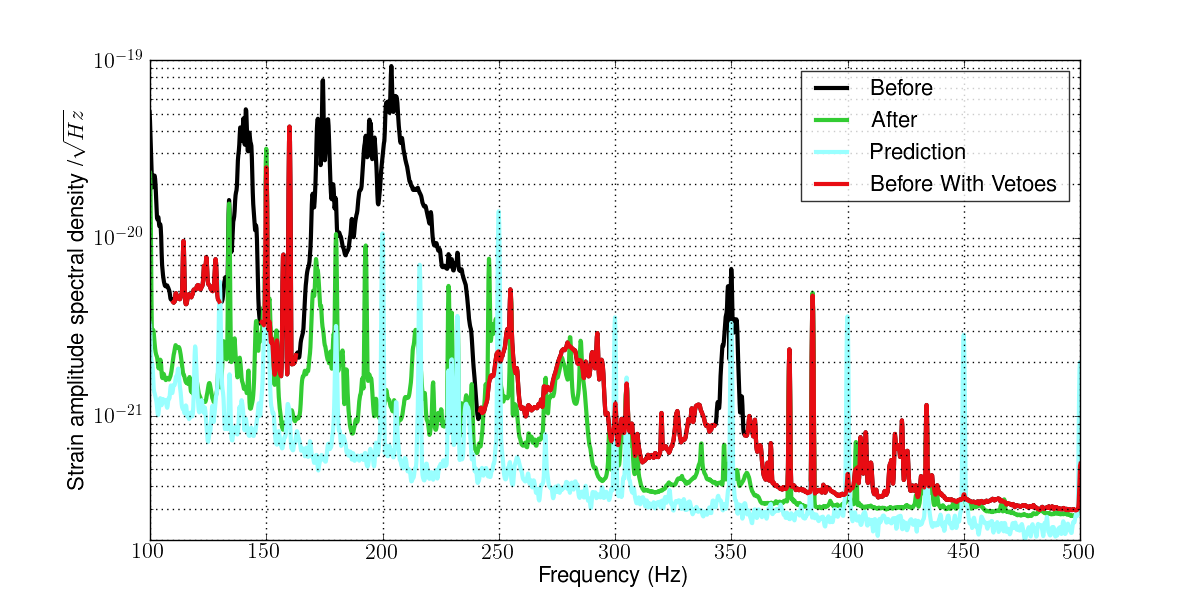}
        \label{fig:bdo_spectrum}
        }
        \subfigure[ Cumulative glitch rate]{\includegraphics[width=0.4\textwidth]{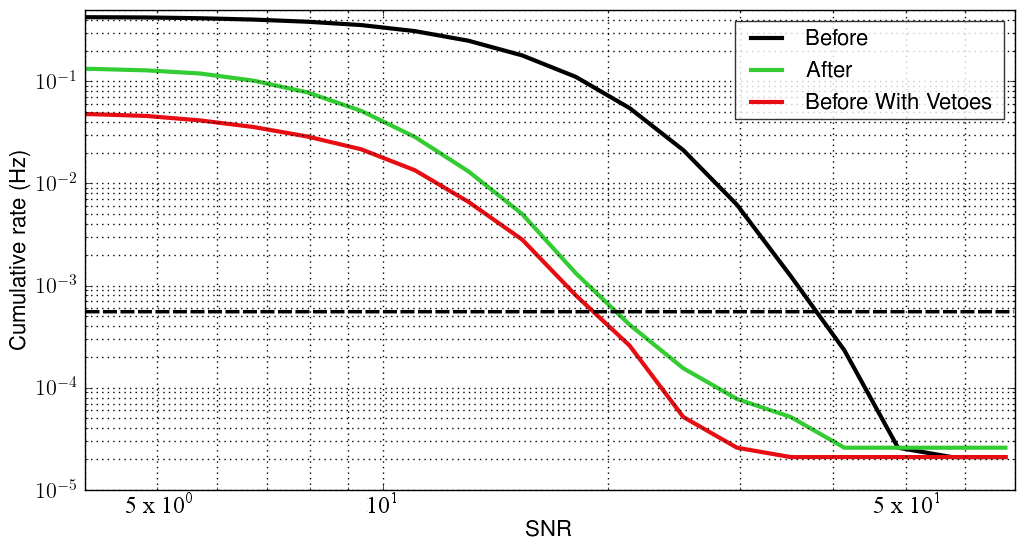}
           \label{fig:bdo_trig_rate}
        }
    \caption{\label{fig:bdo_range_inputs}BDO suspension upgrade comparison plots, in black we show a period of data before the BDO suspension upgrade while a period of data after the upgrade is shown in green.
             The improvement in the noise spectral density from the commissioning can be seen as the reduction of the complex noise features from $100\,\mathrm{Hz}$--$450\,\mathrm{Hz}$ in (a).
             The glitch rate in the $100\,\mathrm{Hz}$--$500\,\mathrm{Hz}$ band with $\mathrm{SNR} \ge 20$ decreased by two orders of magnitude, as shown in (b).
             Also, the SNR threshold, where louder background triggers happen only at a rate lower than $5.6 \times 10^{-4}\,\mathrm{Hz}$ (horizontal dashed black line), is reduced by approximately a factor of two.
             In red we display the effects on these noise characterisations of the application of vetoes defined in the text, while turquoise shows our prediction of the spectral noise reduction from commissioning.}
\end{figure*}
This produced complex noise features in the strain data channel mainly seen from $100\,\mathrm{Hz}$--$500\,\mathrm{Hz}$; see figure \ref{fig:bdo_spectrum}.
When comparing noise spectra from before and after commissioning, we see that at some frequencies there is an increase in the noise by almost a factor of 100.
This noise is also seen very strongly in the cumulative glitch rate histogram shown in figure \ref{fig:bdo_trig_rate}.
For $\mathrm{SNRs} \ge 20$ we see an increase of over a factor 100 in the glitch rate.
More importantly, the SNR threshold at which background triggers happen at a rate of $5.6 \times 10^{-4}\,\mathrm{Hz}$, which is a factor in the computation of the spacetime observation rate, increases by a factor of two.

It should be kept in mind that at the time these effects on the strain data channel were observed, the details of the source of the noise and coupling mechanism were unknown.

\begin{figure}
    \centering
    \includegraphics[width=0.45\textwidth]{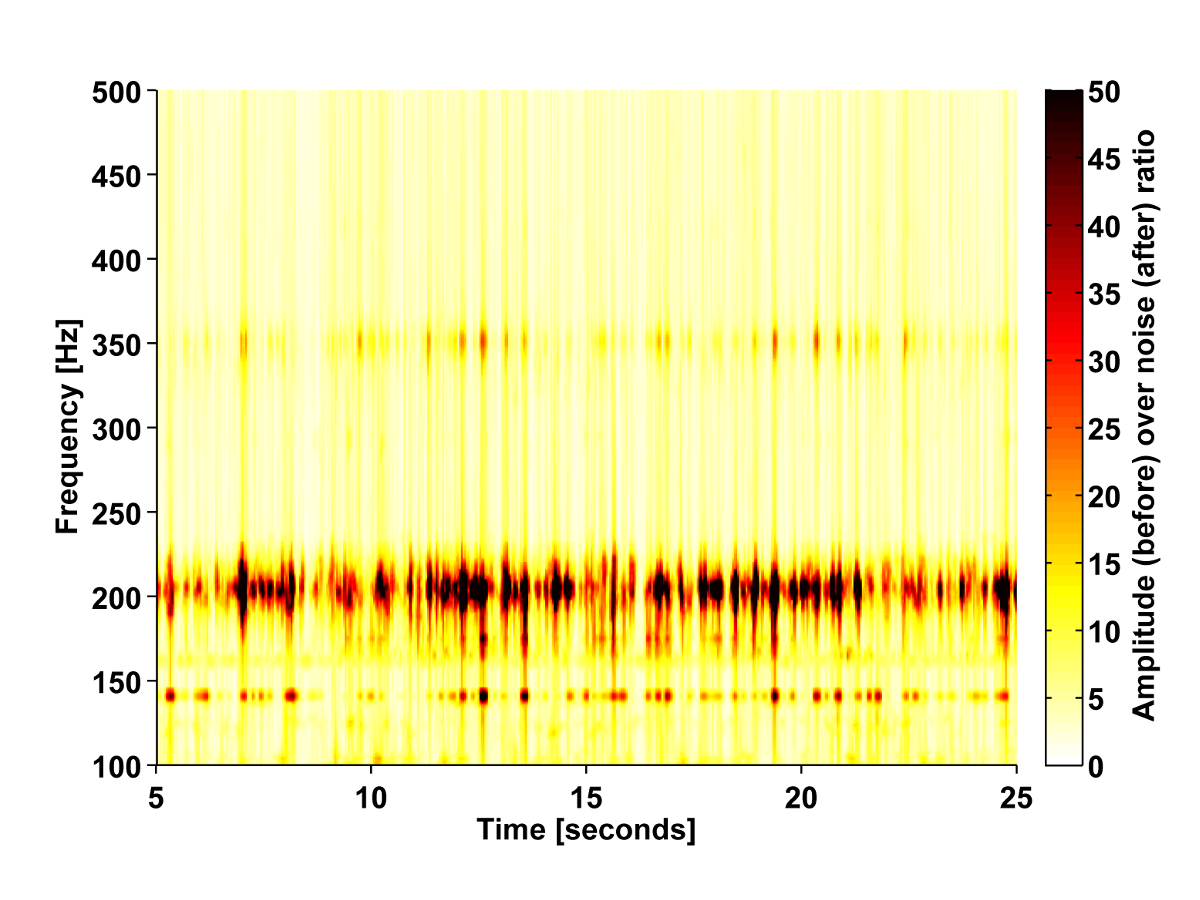}
    \caption{\label{fig:bdo_omegamap}Omegamap showing a period of time before the BDO suspension upgrade.
             Low frequency ($0.5\,\mathrm{Hz}$--$2\,\mathrm{Hz}$) fluctuations in the alignment of the OMC from motion in the interferometer couple BDO suspension jitter in the few hundred Hz range (i.e. $140\,\mathrm{Hz}$) into the detector strain data channel.}
\end{figure}
Plotting the Omegamap of the strain data channel as shown in figure \ref{fig:bdo_omegamap} displays the nature of the noise more clearly, by highlighting two aspects which add to the information provided by the characterisations in figure \ref{fig:bdo_range_inputs}.
The first is that the noise is confined to relatively well-defined frequencies which remain constant.
The second is that the noise is highly variable, at time scales ranging from $0.5$--$2$\,s, with occasional times where all resonant bands (near $140\,\mathrm{Hz}$, $170\,\mathrm{Hz}$, $200\,\mathrm{Hz}$, and $350\,\mathrm{Hz}$) are simultaneously quiet.
With the knowledge of the source of the noise, it is clear where these characteristics come from.
The well defined frequencies are jitter frequencies of the BDO suspensions while the transient nature comes from the motion of the main optic suspensions of GEO\,600.
Here the quiet times are short periods when the OMC passes through the correct alignment and the jitter line coupling is suppressed.

It is already clear from these initial views of the problem how to move towards designing a veto which could be used to mitigate its effects.
We use the well-defined frequency aspect of the noise to create frequency-domain vetoes by notching various frequency bands.
The notched bands from $74\,\mathrm{Hz}$--$110\,\mathrm{Hz}$, $130\,\mathrm{Hz}$--$148\,\mathrm{Hz}$, $163\,\mathrm{Hz}$--$242\,\mathrm{Hz}$ and $343\,\mathrm{Hz}$--$356\,\mathrm{Hz}$ remove $30\%$ of the bandwidth between $100\,\mathrm{Hz}$--$500\,\mathrm{Hz}$.
Vetoing these frequency bands removes the noise source from the data, but also makes the detector blind to GW signals in these frequency bands.
When carrying out an analysis or calculating the spacetime observation rate, we apply these notches to both the strain noise spectral density and the triggers.
For the spectrum, it is assumed that the noise amplitude within the notch band is infinite.
The triggers are notched by removing the glitches with central frequencies within the notched bands.
The effects of this veto on the spectrum and cumulative trigger rate are shown in figure \ref{fig:bdo_spectrum} and \ref{fig:bdo_trig_rate} respectively.
The SNR threshold is reduced by approximately a factor of two.
Table \ref{tab:results} displays the spacetime observation rates for all investigations discussed in this article.
It shows there that this veto increases the spacetime observation rate by a factor of 8.

For this investigation, there was a straightforward way in which a prediction for the effects of commissioning improvements could have been made.
This noise was seen on switching the readout technique from a heterodyne-readout to a DC-readout \cite{Affeldt:2014ge}.
Therefore it could be assumed that with commissioning it would be possible to reduce the noise back down to the level seen with GEO\,600 configured in the heterodyne-readout scheme.
Thus we take one of the heterodyne-readout strain noise spectra in the frequency band of interest as a prediction of the commissioning improvements.
This is shown in figure \ref{fig:bdo_spectrum}.
We combine this spectrum with the noise transients from the before commissioning period with the frequency vetoes applied to calculate the predicted range.
This gives us a predicted burst-like spacetime observation rate of $300 \times 10^5\,\textrm{kpc}^{3}$ as can be seen in table \ref{tab:results}---a factor of 50 improvement over the before-commissioning observation rate and a factor 7 over the with-vetoes rate.

\begin{figure}
    \centering
    \includegraphics[width=0.45\textwidth]{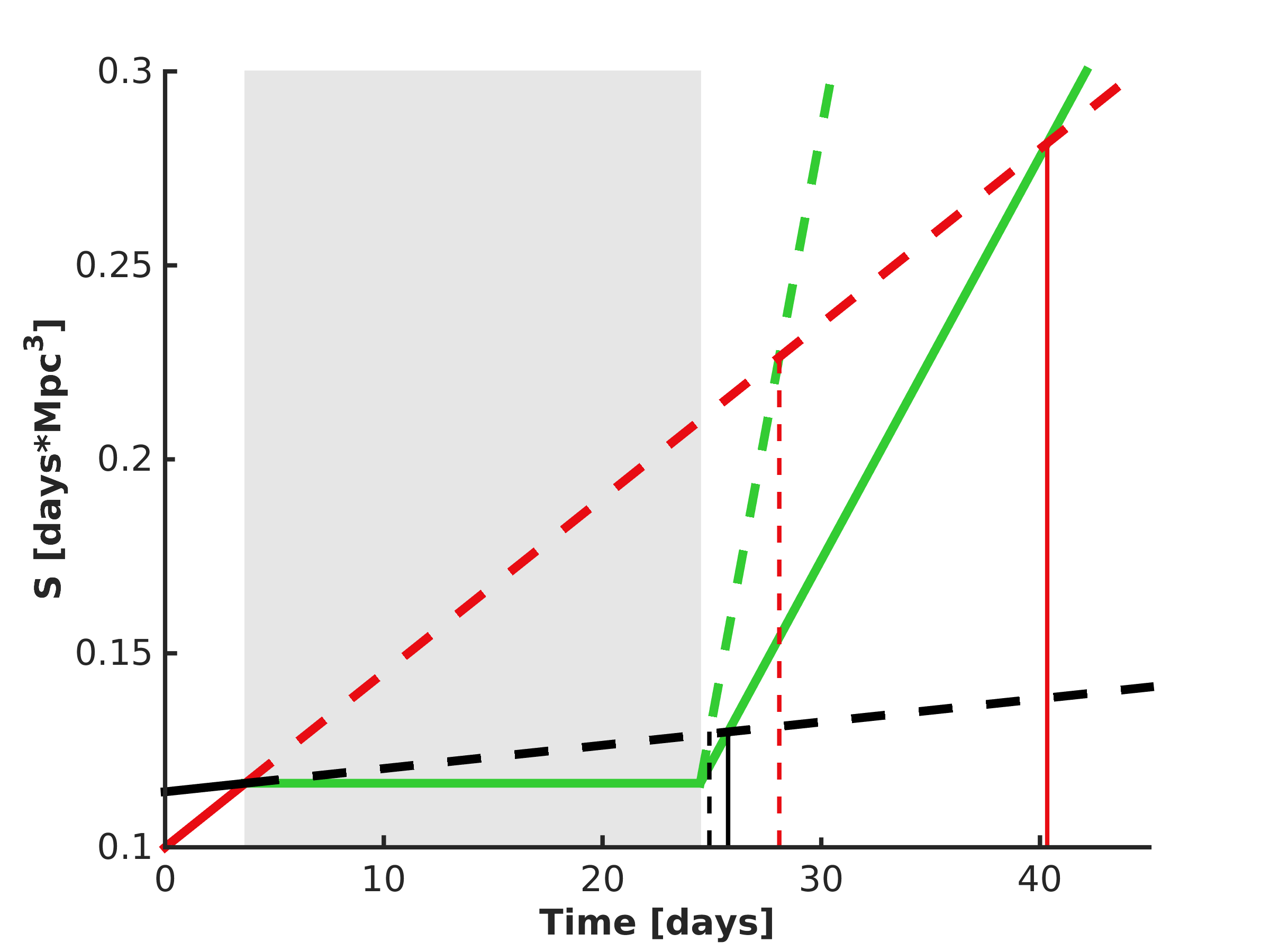}
    \caption{\label{fig:bdo_commvis}Graphical display of the spacetime observation volume for a GW search, $S \equiv \int \dot{S} \, \mathrm{d} t$ around the time of the BDO suspension upgrade.
      The period displayed here starts from sometime before the commissioning period until after all benefit-delay times have been reached.
      All spacetime observation volumes are shifted so they are equal at the beginning of the commissioning period which is designated by the grey filled in area.
      Here all real spacetime observation volumes are plotted as solid lines while extrapolated or predicted spacetime observation volumes are plotted as dashed lines.
      In black we plot quantities which correspond to the before-commissioning period without vetoes while red denotes those quantities with vetoes.
      The green lines plot spacetime observation volumes during and after commissioning (hypothetical and real).
      The benefit-delay times are displayed by the vertical lines where the hypothetical continued before- and veto-spacetime observation volume meets the after- or predicted-spacetime observation volume.}
\end{figure}
With vetoes and a prediction for improvements from commissioning, we can carry out a benefit-delay analysis as described in section \ref{sec:tools}.
The results are given in table \ref{tab:benefit-delay}, as well as in figure \ref{fig:bdo_commvis} where we have already applied the actual amount of time which was needed for commissioning the improvements.
Here we see that although the veto already provides very substantial improvements to the spacetime observation rate, the prediction promises such large reductions in noise that the astrophysical losses from commissioning is made up within a fraction of the time taken to do the commissioning.
Since at this time we knew that GEO\,600 would be running for at least three full years more, commissioning an improvement would have to take more than $2.5$ years of lost operation time for it to be no longer worthwhile.
In this case it would have been clear that commissioning interruptions should proceed.

From the spectral characteristics of the noise alone, however, invasive commissioning investigations were already deemed necessary before seeing such benefit-delay time analyses although it was unclear how much time would need to be devoted to find the source of the noise.
Once the source was identified as described in the beginning of this section, the BDO suspensions were redesigned to provide roughly $100$ times better isolation of the ground motion from $50\,\mathrm{Hz}$--$1\,\mathrm{kHz}$.
This reduced the high frequency fluctuations which were being modulated to create the large noise structures in figure \ref{fig:bdo_spectrum}.
The cumulative down-time from the investigations as well as commissioning of this upgrade was $20.8$\,days.
The effect of this commissioning on the spectrum and cumulative trigger rate are shown in figure \ref{fig:bdo_spectrum} and \ref{fig:bdo_trig_rate} respectively.
The transient noise shows improvements very similar to the improvements gained by applying the vetoes.
For the spectrum we also observe large gains compared to the noise level before commissioning.
Firstly, the noise is reduced over almost the entire band.
Secondly, many of the large noise features are significantly reduced: in the best case, near $200\,\mathrm{Hz}$, by almost a factor of $100$.
These improvements together contribute to a 17-fold increase in the spacetime observation rate over the before-commissioning rate and only a 2-fold increase for the vetoed-case.

Using these actual results of the commissioning improvements we can again carry out a benefit-delay time analysis to see if the commissioning interruptions were actually worthwhile.
In table \ref{tab:benefit-delay} and figure \ref{fig:bdo_commvis} we see that although the improvement turned out not to be better than the prediction, the deficit in potentially observed sources caused by the commissioning is very quickly made up by the improvements seen.
The benefit-delay ratio for the actual improvements, with vetoes applied, is $0.75$ so we would have always recovered from the number-of-sources deficit in less time than the total commissioning down-time.
Through our cost-benefit analysis we see that the commissioning interruption was worthwhile.
Although this was the assumed outcome simply by looking at the spectral noise amplitude, the quantitative nature of this analysis justifies this assumption.
In the following examples, we will see cases where these benefit-delay time analyses would have had an impact on the way we thought about the commissioning---either because of the subtlety of the problem or the power of the vetoes.

\begin{figure*}
    \centering
        \subfigure[ Strain noise spectral density]{\includegraphics[width=0.45\textwidth]{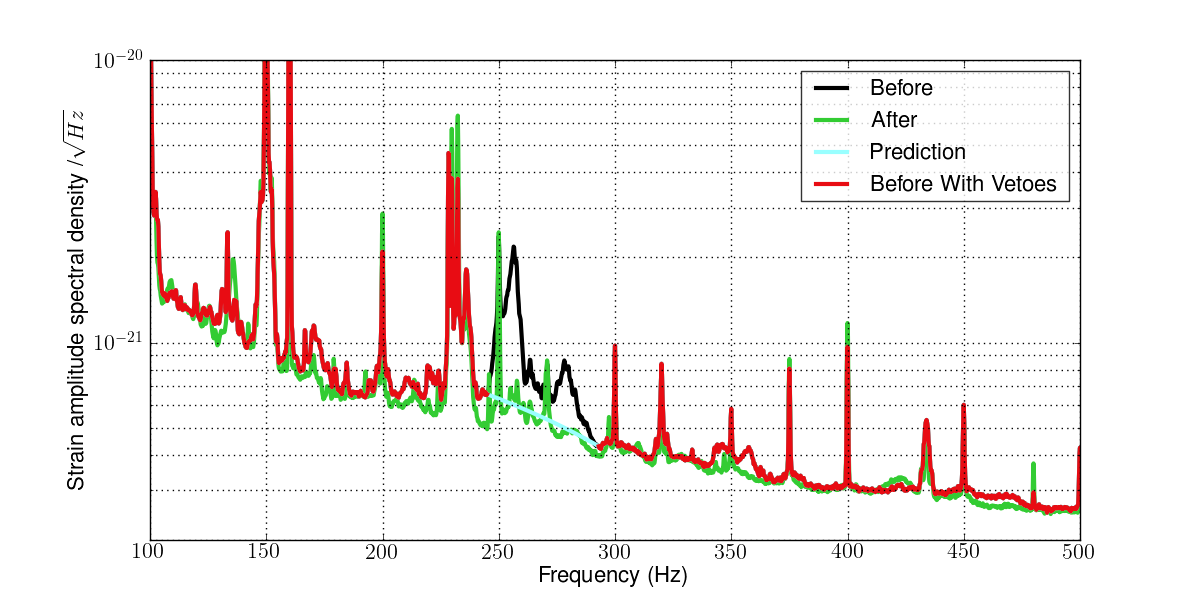}
        \label{fig:omc_isolation_spectrum}
        }
        \subfigure[ Cumulative glitch rate]{\includegraphics[width=0.4\textwidth]{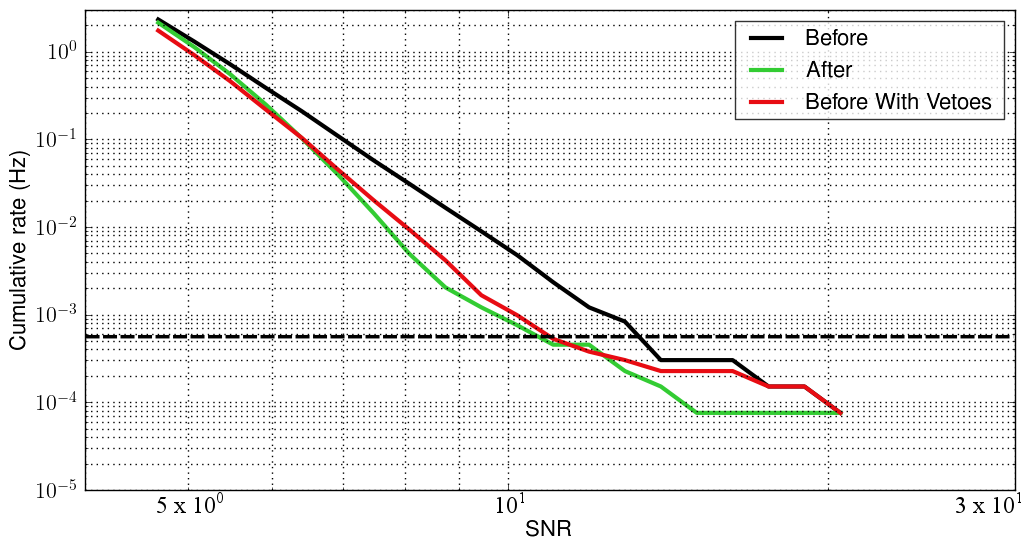}
        \label{fig:omc_isolation_trig_rate}
        }
    \caption{\label{fig:omc_isolation_range_inputs}OMC isolation upgrade comparison plots, in black we show a period of data before the OMC isolation upgrade while a period of data after the upgrade is shown in green.
             The noise features are well confined to two narrow frequency bands around $255\,\mathrm{Hz}$ and $280\,\mathrm{Hz}$, where we can easily select a notch veto (red) to remove these features from the spectrum, as can be seen in (a).
             The glitch rates after applying vetoes and after commissioning are similar: both are about a factor 5 lower than the before-commissioning level for $\mathrm{SNRs} \ge 20$ as can be seen in (b).
             However, the SNR threshold, where louder background triggers happen only at a rate lower than $5.6 \times 10^{-4}\,\mathrm{Hz}$ (horizontal dashed black line), is reduced only by 20\%.
             The turquoise curve in (a) shows our prediction of the spectral noise reduction from commissioning.}
\end{figure*}

We take a look now at a noise source involving the same mechanism as the noise caused by the BDO suspensions but with far subtler effects on the strain data.
This source has resonances of elements within the OMC and its suspension.
The resultant noise had been previously buried by the noise caused by the BDO suspensions so was only noticed after the BDO suspension upgrade.
Here we observed resonant structures near $255\,\mathrm{Hz}$ and $280\,\mathrm{Hz}$ which can be seen by comparing the before- and after-commissioning spectra in figure \ref{fig:omc_isolation_spectrum}.
These structures are not more than a factor $4$ above the underlying noise floor.
In the cumulative glitch rate histogram (figure \ref{fig:omc_isolation_trig_rate}), triggers with $\mathrm{SNR} \ge 10$ occur a little less than $8$ times more often when the noise is present than when it is not present and the SNR threshold is about 30\% higher.

\begin{figure}
    \centering
    \includegraphics[width=0.45\textwidth]{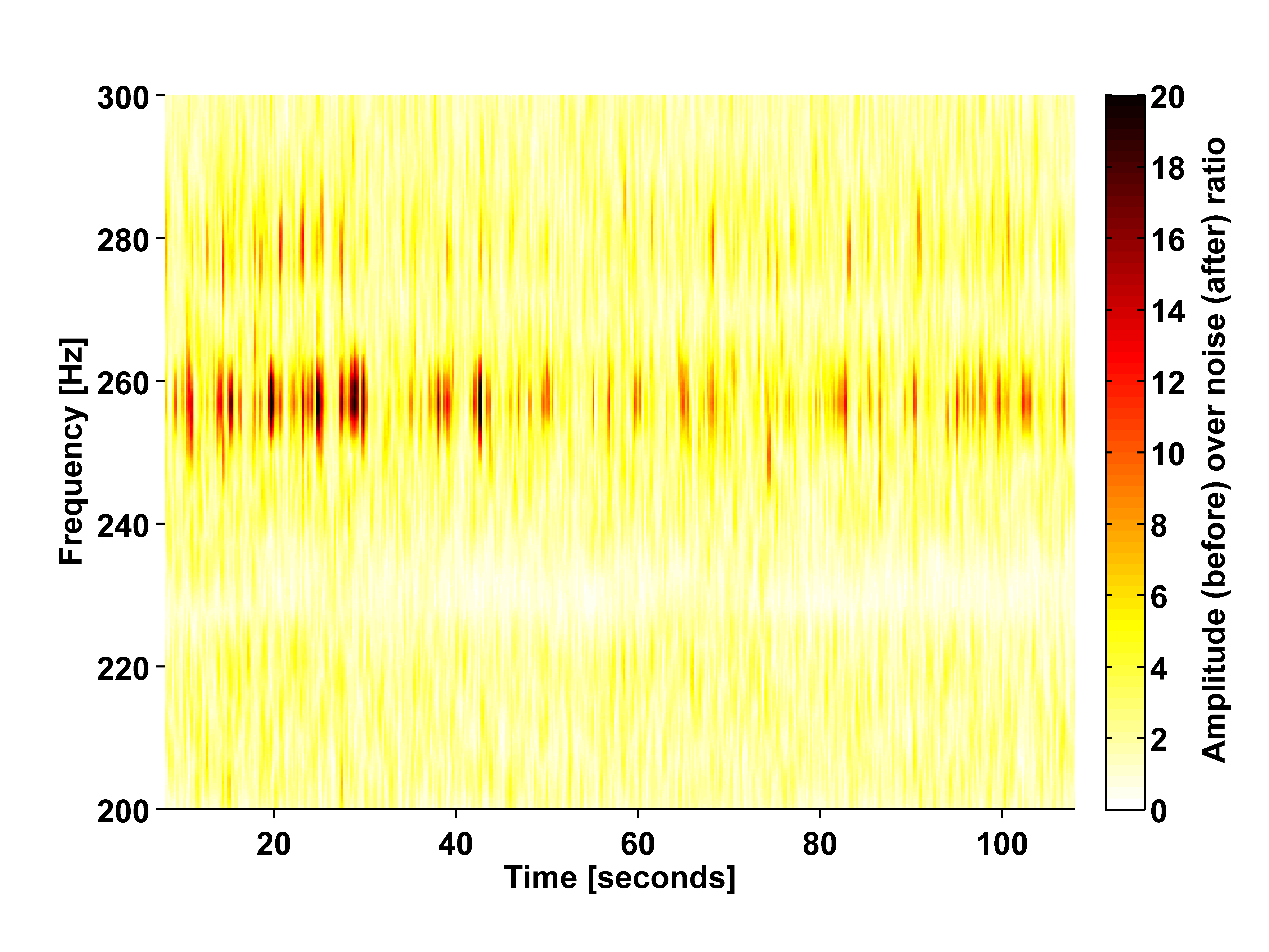}
    \caption{\label{fig:omc_isolation_omegamap}Omegamap showing a period of time with transient noise in the two bands around $255\,\mathrm{Hz}$ and $280\,\mathrm{Hz}$.
             Fluctuations in the alignment of the OMC allow noise entering the detector through the seismic isolation of the OMC to be coupled into the strain data channel.}
\end{figure}
The nature of the noise can be seen in the Omegamap of the detector output channel in figure \ref{fig:omc_isolation_omegamap}.
Comparing the Omegamaps figure \ref{fig:bdo_omegamap} and figure \ref{fig:omc_isolation_omegamap}, without previous knowledge of the mechanisms, we see that these noise sources have similar frequency characteristics (both stable) and similar transient behaviour.
From these plots one may already guess that the mechanism for coupling of motion into the strain data channel for both noise scenarios is the same.

As before, a frequency dependent veto can notch out the glitch bands while maintaining sensitivity at other frequencies.
The veto used here is a notch at $243\,\mathrm{Hz}$--$293\,\mathrm{Hz}$ applied in the same manner as done above.
This time the veto only removes $13$\% of the bandwidth between $100\,\mathrm{Hz}$--$500\,\mathrm{Hz}$.
The effects of this veto on the spectrum and cumulative trigger rate are shown in figure \ref{fig:omc_isolation_spectrum} \ref{fig:omc_isolation_trig_rate} respectively.
The SNR threshold is only reduced by 16\%.
In table \ref{tab:results} we see that the spacetime observation rate increases by 60\% through the use of the veto.

With the hint from the Omegamap that this noise originates from a mechanism similar to that of the BDO suspension noise, it is not difficult to come up with a prediction on what we can achieve with commissioning.
Here, since only a factor of $5$ is needed, we assume that we can suppress these resonant structures to a level where they would lie below the underlying noise floor.
We then take as a predicted spectrum, one where we have simply interpolated across the notch as shown in figure \ref{fig:omc_isolation_spectrum}.
Combining this spectrum with the noise transients from the before-commissioning period, with the frequency veto applied, then allows us to calculate the predicted range.
This gives us a predicted burst-like spacetime observation rate of $850 \times 10^5\,\textrm{kpc}^{3}$ as can be seen in table \ref{tab:results}---an 80\% improvement over the before-commissioning spacetime observation rate but only a 13\% improvement over the with-vetoes rate.

\begin{figure}
    \centering
    \includegraphics[width=0.45\textwidth]{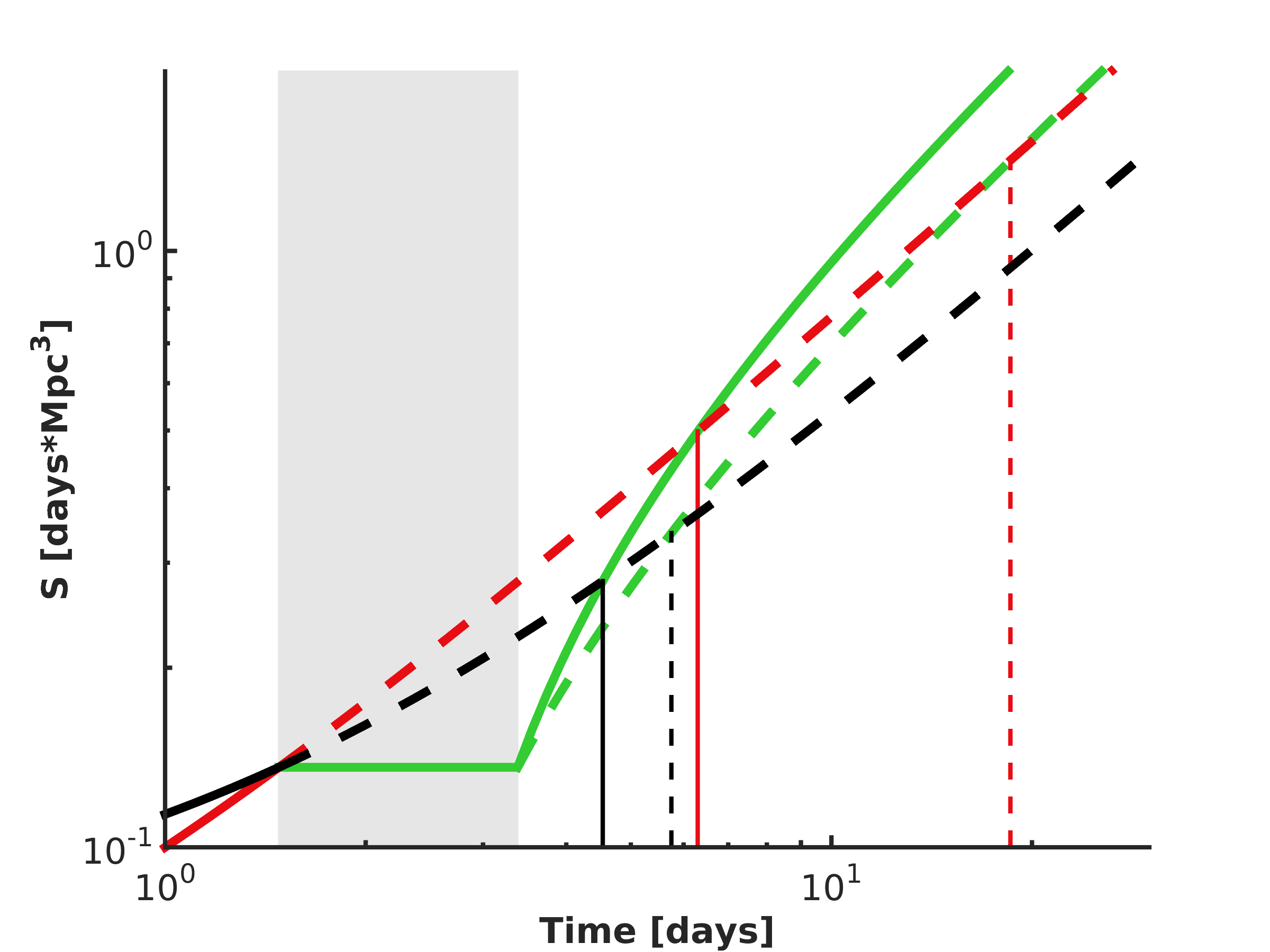}
    \caption{\label{fig:omc_commvis}Graphical display of the spacetime observation volume for a GW search, $S \equiv \int \dot{S} \, \mathrm{d} t$ around the time of the OMC isolation upgrade.
      The period displayed here starts from sometime before the commissioning period until after all benefit-delay times have been reached.
      All spacetime observation volumes are shifted so they are equal at the beginning of the commissioning period which is designated by the grey filled in area.
      Here all real spacetime observation volumes are plotted as solid lines while extrapolated or predicted spacetime observation volumes are plotted as dashed lines.
      In black we plot quantities which correspond to the before-commissioning period without vetoes while red denotes those quantities with vetoes.
      The green lines plot spacetime observation volumes during and after commissioning (hypothetical and real).
      The benefit-delay times are displayed by the vertical lines where the hypothetical continued before- and veto-spacetime observation volume meets the after- or predicted-spacetime observation volume.}
\end{figure}
Carrying out a benefit-delay time analysis using the spacetime observation rates from the vetoed- and predicted-scenario then yields the results in table \ref{tab:benefit-delay}, figure \ref{fig:omc_commvis} where we get a benefit-delay ratio of $8$.
This ratio is much larger than for the noise which called for the BDO suspension upgrade, however, still low enough that commissioning interruptions for investigations and improvements would be beneficial.
Even if the commissioning down-time would have taken as long as the BDO suspension upgrade down-time, the deficit in the potentially observed GW sources would be made up within a half year of observation.
However, due to observations of a change in the resonant noise structures after an instrumental change commissioned for other reasons, a little after this problem was observed it was already clear that a solution would require much less time.
The commissioning interruption was thought to take no more than $3$\,days which would bring the benefit-delay time down to $3.5$\,weeks.
Given these considerations, it is clear that commissioning a solution to this noise should proceed as soon as possible.
Due to the subtle nature of the problem, without the benefit-delay time analysis this conclusion is not obvious.

The serendipitous observation of a change in the two resonant structures happened directly after work on the OMC optical bench.
Thus we decided to improve the isolation of this bench from ground motion as an attempt to remove these noise structures.
The cumulative down-time from the required investigations and commissioning work was $1.9$\,days.
The effect of this commissioning on the spectrum and cumulative trigger rate are shown in figure \ref{fig:omc_isolation_spectrum} and \ref{fig:omc_isolation_trig_rate} respectively.
The amplitude of the resonance structures at $255\,\mathrm{Hz}$ and $280\,\mathrm{Hz}$ are reduced by a factor of almost $3$ and $2$ respectively, see figure \ref{fig:omc_isolation_spectrum}, such that they are beneath the noise floor.
We also see small improvements in other features in the noise floor across the $100\,\mathrm{Hz}$--$500\,\mathrm{Hz}$ band.
Similar but a little more effective than the application of the veto, SNR threshold reduces by a factor of $22\%$, as seen in figure \ref{fig:omc_isolation_trig_rate}.
These small improvements lead to a 3-fold increase in the spacetime observation rate over the before-commissioning rate and a $70\%$ increase over the vetoed-case.
Comparing these improvements to the predicted values in table \ref{tab:results} shows that we have, this time, made conservative predictions.

Putting the actual spacetime observation rate into the benefit-delay time analysis then highlights the fact that the commissioning brought better improvements than predicted by lowering the benefit-delay ratio from $8$ to $1.5$, as shown in table \ref{tab:benefit-delay} and graphically in figure \ref{fig:omc_commvis}.
This shows that the decision to proceed with commissioning was in fact well justified.

In contrast to the noise-case solved by upgrading the BDO suspensions, we show that the decision to commission an upgrade here would have attracted more weight had a benefit-delay time analysis been carried out.
In the next section we will look at an example for which a benefit-delay time analysis gives key insight into the best solution for a severe transient noise source.

\subsection{Squeezer glitches}
\label{sec:squeezer}

\begin{figure}
    \centering
    \includegraphics[width=0.45\textwidth]{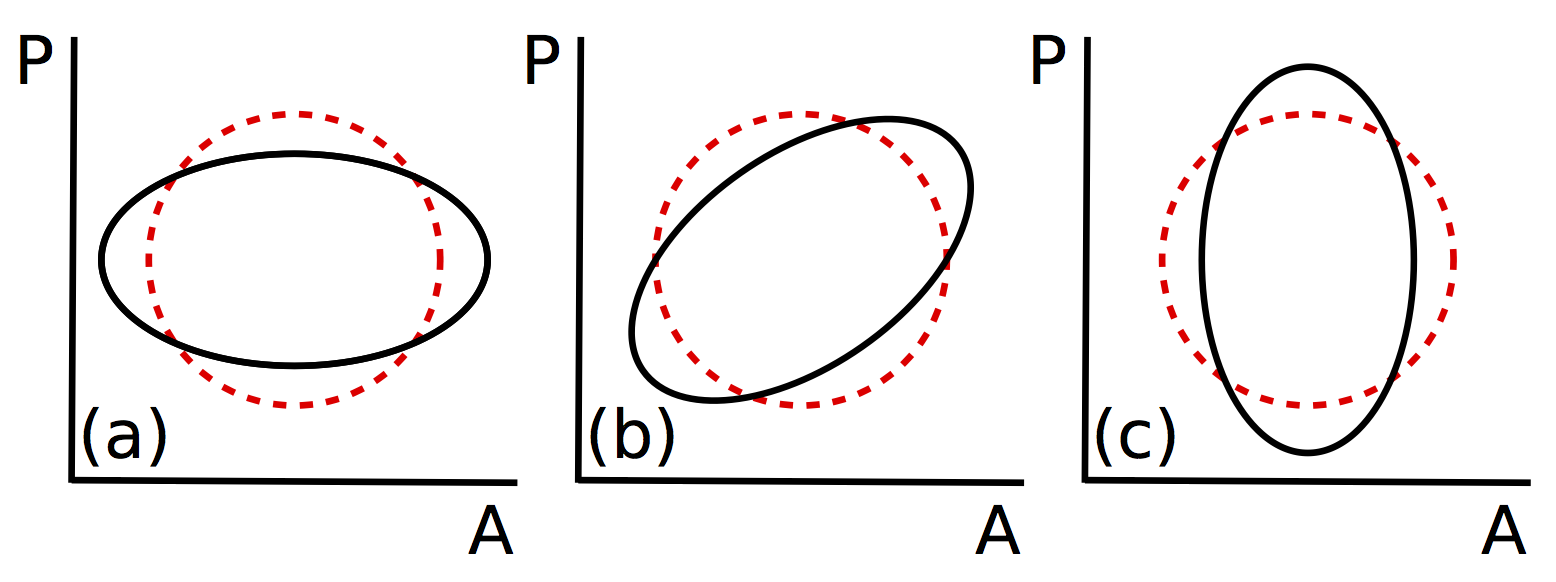}
    \caption{\label{fig:sqz_ellipse}In the phase ($P$) and amplitude ($A$) space of the light field inside an interferometer we show the error ellipses for quantum fluctuations of a coherent vacuum state (dashed lines) compared to those of squeezed states (solid lines).
            The three plots show different example directions in which quantum fluctuations can be suppressed by squeezing.
            A Michelson interferometer makes measurements in the phase quadrature of the light field so that we can represent the level of shot noise of the beam at the output of the interferometer schematically by the width of the ellipse in the $P$ direction.
            In the region where the interferometer's sensitivity is limited by shot noise the best sensitivity improvement is then achieved by squeezing the ellipse in the $P$ direction as shown in (a).
            (b) shows squeezing along a direction which gives the same shot noise as the coherent vacuum state, i.e. where there is no effect from the squeezing.
            (c) shows squeezing along the direction which increases the shot noise level in the interferometer output and thus decreases its sensitivity to GWs.}
\end{figure}

\begin{figure*}
    \centering
        \subfigure[ Strain noise spectral density]{\includegraphics[width=0.45\textwidth]{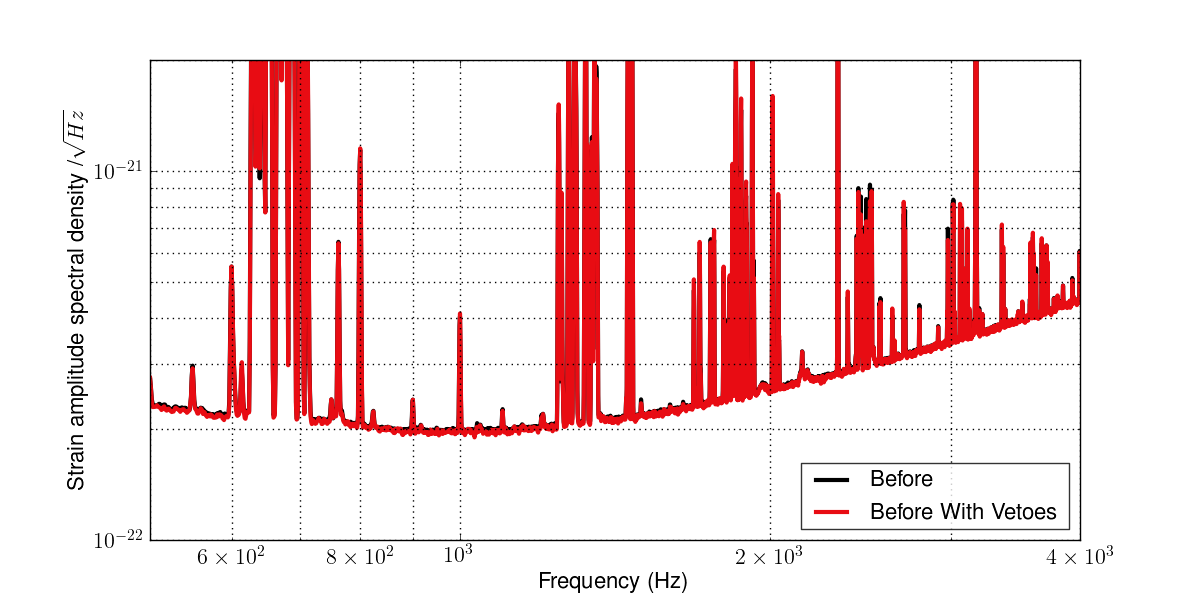}
        \label{fig:sqz_ep_spectrum}
        }
        \subfigure[ Cumulative glitch rate]{\includegraphics[width=0.4\textwidth]{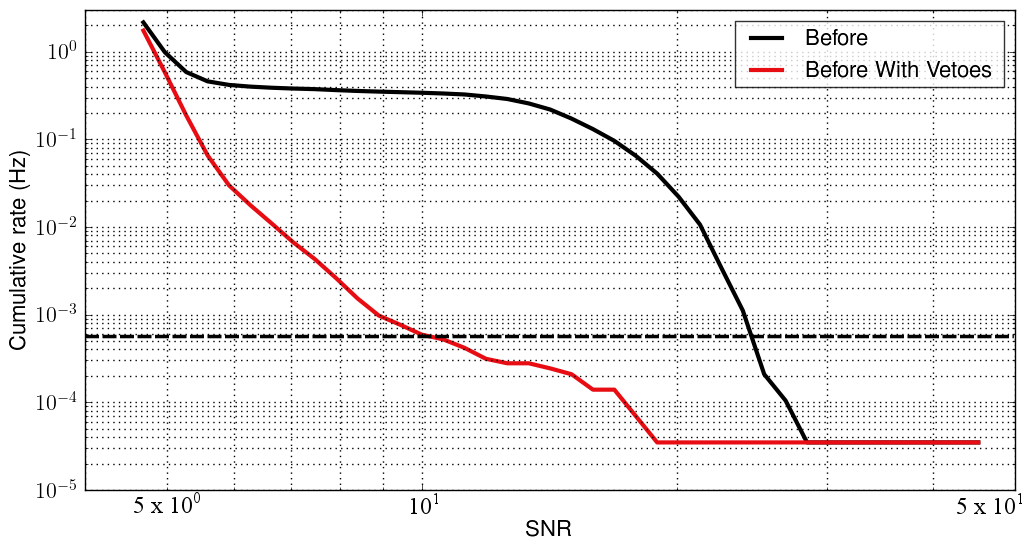}
        \label{fig:sqz_ep_trig_rate}
        }
    \caption{Squeezer glitches comparison plots.
             A period of time when glitches caused by the squeezer were happening is shown in black, while the same period with vetoes applied is shown in red.
             (a) The transient noise events are loud, but they do not happen frequently enough to make a significant change to the strain noise spectral density, when averaged over the period of data used for this investigation.
             (b) Both the vetoes and commissioning do well to remove the large number of high SNR glitches that were being produce by this noise source.
             The horizontal dashed black line shows the false-alarm rate threshold of $5.6\times10^{-4}\,\mathrm{Hz}$.}
\end{figure*}

\begin{figure}
    \centering
    \includegraphics[width=0.45\textwidth]{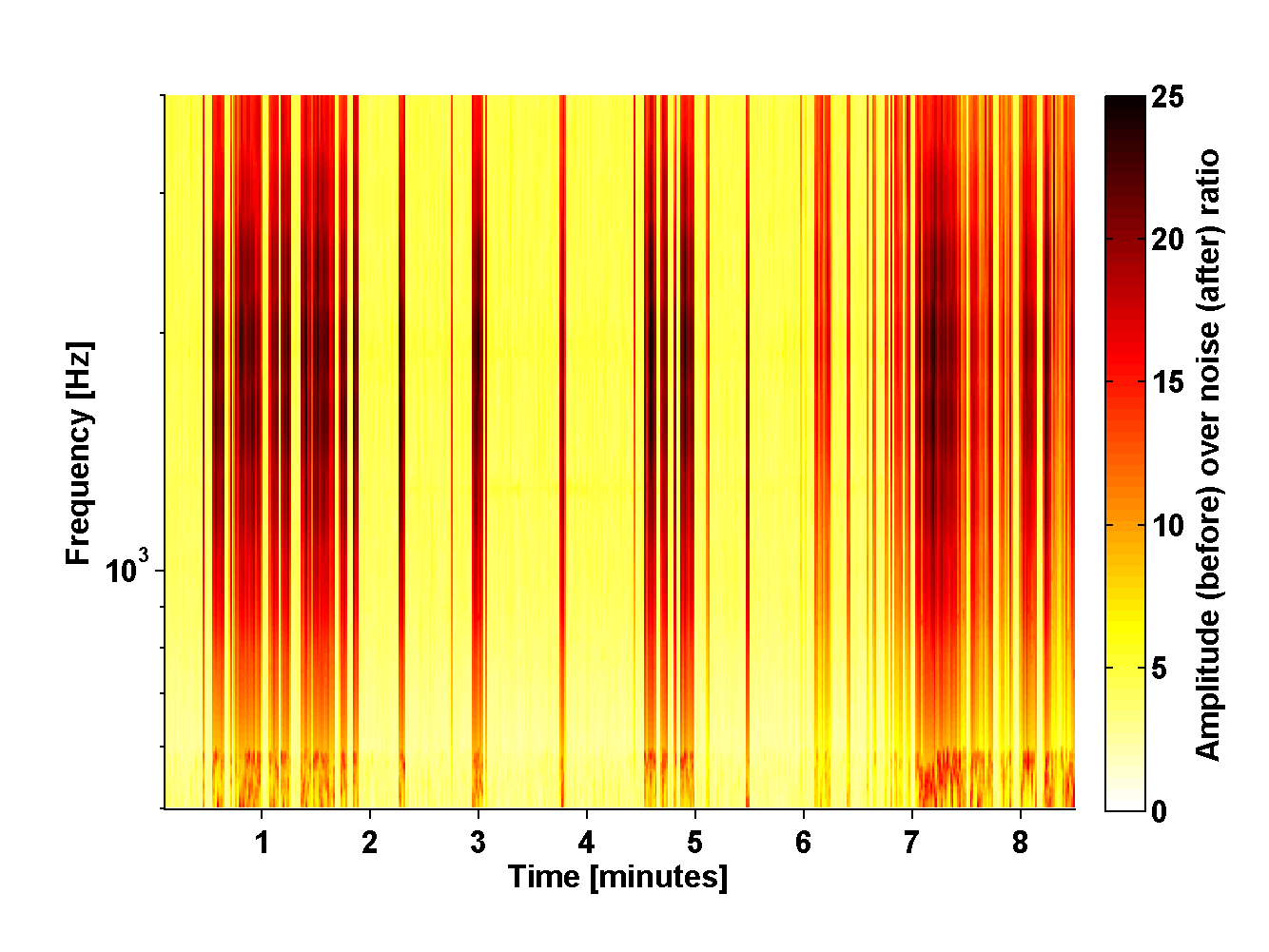}
    \caption{\label{fig:sqz_ep_omegamap}Omegamap showing the transient noise events caused by the squeezer.}
\end{figure}

\begin{figure}
    \centering
    \includegraphics[width=0.45\textwidth]{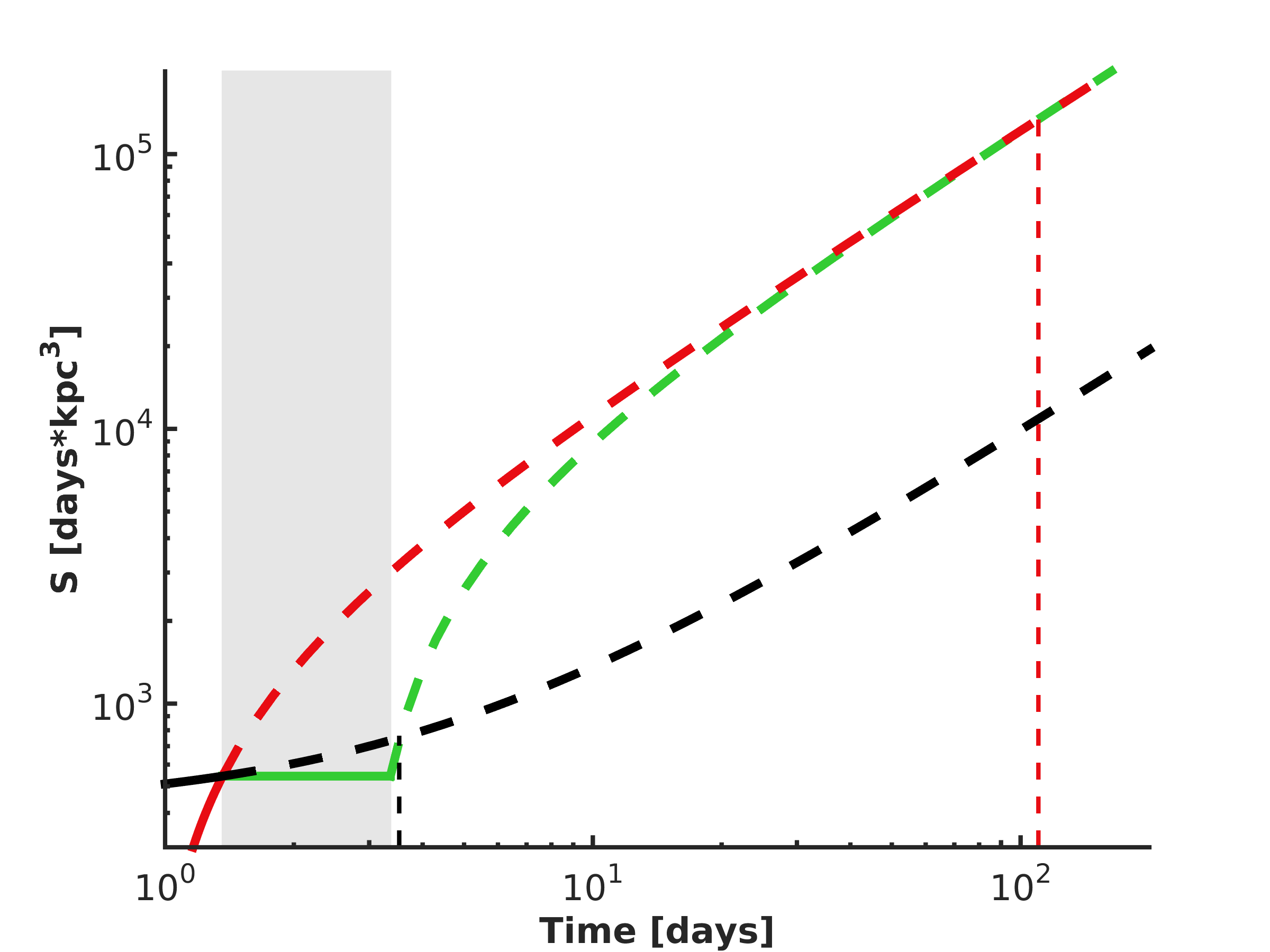}
    \caption{\label{fig:sqzep_commvis}Graphical display of the spacetime observation volume for a GW search, $S \equiv \int \dot{S} \, \mathrm{d} t$ around the time of a hypothetical 2\,day commissioning interruption aimed at removing the squeezer glitches.
      The period displayed here starts from sometime before the commissioning period until after all benefit-delay times have been reached.
      All spacetime observation volumes are shifted so they are equal at the beginning of the commissioning period which is designated by the grey filled in area.
      Here all real spacetime observation volumes are plotted as solid lines while extrapolated or predicted spacetime observation volumes are plotted as dashed lines.
      In black we plot quantities which correspond to the before-commissioning period without vetoes while red denotes those quantities with vetoes.
      The green lines plot spacetime observation volumes during and after commissioning (hypothetical).
      The benefit-delay times are displayed by the vertical lines where the hypothetical continued before- and veto-spacetime observation rate meets the predicted-spacetime observation volume.}
\end{figure}

GEO\,600 is the first large scale interferometric detector to implement the injection of squeezed light to improve the shot noise limited sensitivity \cite{Abadie:2011dj}, and also the first such detector to operate this technique on year time scales \cite{Grote:2013sq} and during science operations \cite{Aasi:2014ga}.

To gain an understanding of squeezing, we can picture the quantum uncertainty by drawing an ellipse in a plane spanned by two conjugate variables, which for laser light are the phase and amplitude of the electric field as shown in figure \ref{fig:sqz_ellipse}.
The minimum area of this ellipse is set by the Heisenberg uncertainty principle.
Squeezed states of light are quantum states that have a reduced uncertainty in one of the field quadratures compared to a coherent state, and an increased uncertainty in the conjugate quadrature.
A squeezed state of light is injected into the output port of the interferometer (see figure \ref{fig:geo_layout}) to replace the coherent vacuum state that would otherwise be entering this port.
If the squeezing is oriented correctly in relation to the electric field inside the interferometer it will improve the shot noise limited sensitivity of the interferometer \cite{Caves:1981hw}; see figure \ref{fig:sqz_ellipse}.
The orientation of the squeezing error ellipse with relation to the electric field inside the interferometer is controlled to optimise the improvement for the shot noise limited sensitivity as shown in figure \ref{fig:sqz_ellipse}.

With GEO\,600 leading the way in the implementation of squeezed light we have come across previously unseen noise sources.
A population of transient noise events, ``glitches'', was observed that produced broadband noise in the strain data channel.
The noise is short duration and broadband above a few hundred Hz.
An example of this can be seen in figure \ref{fig:sqz_ep_omegamap}.
These glitches only occur during periods when the squeezer is in operation.

It was found that these glitches were due to the squeezer malfunctioning and producing noise that was being injected into the interferometer, caused by noise in the squeezer ellipse orientation control loop.
Hence the error point of this loop was a viable witness channel for this noise source, and was used to produce vetoes to remove the data contaminated by these glitches.
An investigation correlated the low-passed time series of the squeezer error point to the amplitude and rate of glitches in the strain data channel.
A threshold on the voltage of this channel was used to produce vetoes which removed the majority of the glitches, with high efficiency.
These vetoes incur a $1.85\%$ down-time during science operations.
The vetoes generated removed time intervals without frequency dependence as a result of the broadband nature of the noise contamination.

The effect of this veto on the spectrum and cumulative glitch rate are shown in figure \ref{fig:sqz_ep_spectrum} and \ref{fig:sqz_ep_trig_rate} respectively.
The strain noise spectral density is not obviously effected---although the transient noise events are large in amplitude they do not happen frequently enough to make a significant contribution when averaging over the period of data used for this investigation.
In the cumulative glitch rate histogram, the removal of these glitches causes the SNR threshold to be reduced by a factor of $2.5$.
In table \ref{tab:results} we see that the vetoes, by removing this noise source, improve the burst-like spacetime observation rate from $100\,\textrm{kpc}^{3}$ to $1200\,\textrm{kpc}^{3}$.

Although we did not obtain a full commissioning solution for this problem, we can make a prediction of the performance of the detector with this noise source removed.
To predict the spacetime observation rate without this noise source present, we use the data with our veto applied but omit the $1.85\%$ down-time it incurs.
This yields a spacetime observation rate of $1300\,\textrm{kpc}^{3}$.

Using these values of the spacetime observation rate we find that the predicted benefit-delay ratio from removing this noise source completely is $0.08$.
However when first applying the veto, the predicted benefit-delay ratio from removing the noise source completely is $53$.
This increase shows that the veto performs very well and in many situations would make commissioning not beneficial to the observing capabilities of a science run.

Without performing our cost-benefit analysis, the severity of this noise source as seen in figure \ref{fig:sqz_ep_omegamap} and \ref{fig:sqz_ep_trig_rate} could encourage a rash decision to perform commissioning and investigative work to attempt to locate and remove the source of this noise.
However, after applying our cost-benefit analysis we see that any down-time induced by such work would produce a deficit in our observable spacetime volume which would only be compensated by the benefits of a fix in a period of time 53 times as long as the down-time.
Since it was not at all clear how long of a down-time was needed to investigate the problem and commission a solution, any investigations which would carry on for more than a few days without successfully finding the source of the noise would start to be in danger of never being able to collect the benefits.

In this example, our cost-benefit analysis has allowed us to quantitatively determine if a noise problem is occurring often enough to warrant commissioning.
The infrequency of the squeezer glitches and the existence of a good veto turned a severe problem into one that is not worth taking the observation time to investigate.
For our next and last example, we describe a case where ideas similar to the ones presented in this article were applied in a decision-making process which ultimately had an impact to the commissioning actions.

\subsection{$3.5\,\mathrm{Hz}$ dither squeezer glitches}
\label{sec:dither_glitches}

A second class of glitches associated with the squeezer were observed during the S6E/VSR4 science run in 2011.
Sinusoidal modulations are applied to two of the three BDOs to two angular degrees of freedom per optic.
Each of these four dithers are at different frequencies.
The purpose of the dithers is to create error signals used to automatically maintain the alignment of the interferometer beam to the output mode cleaner.
A $3.5\,\mathrm{Hz}$ dither, specifically, modulates the degree of freedom with the largest geometric effect on the squeezer path.

\begin{figure*}
    \centering
        \subfigure[ Cumulative glitch rate for triggers above $1\,\mathrm{kHz}$]{\includegraphics[width=0.4\textwidth]{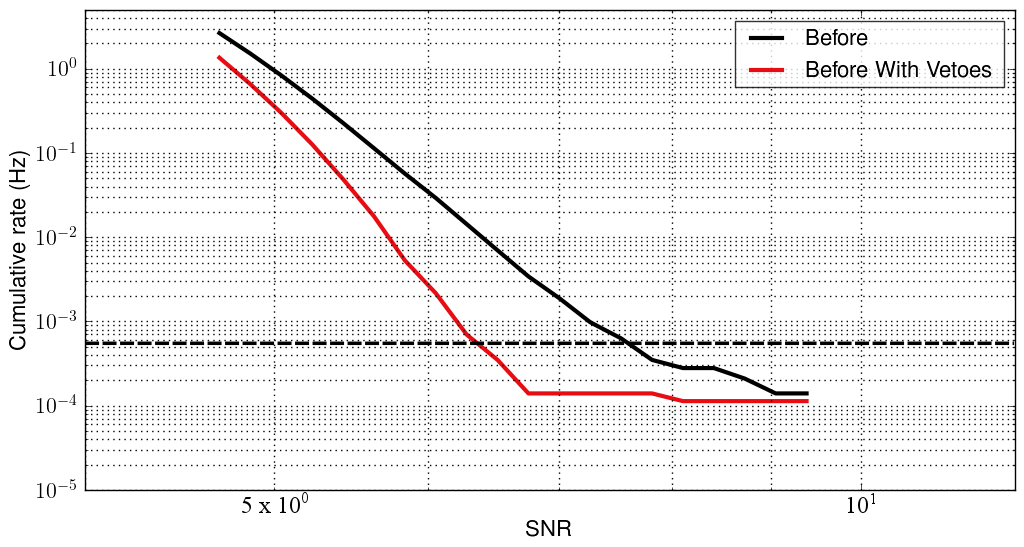}
        \label{fig:sqz_backscatter_trig_rate-band}
        }
        \subfigure[ Histogram of Omega trigger time modulo $2$ seconds]{\includegraphics[width=0.45\textwidth]{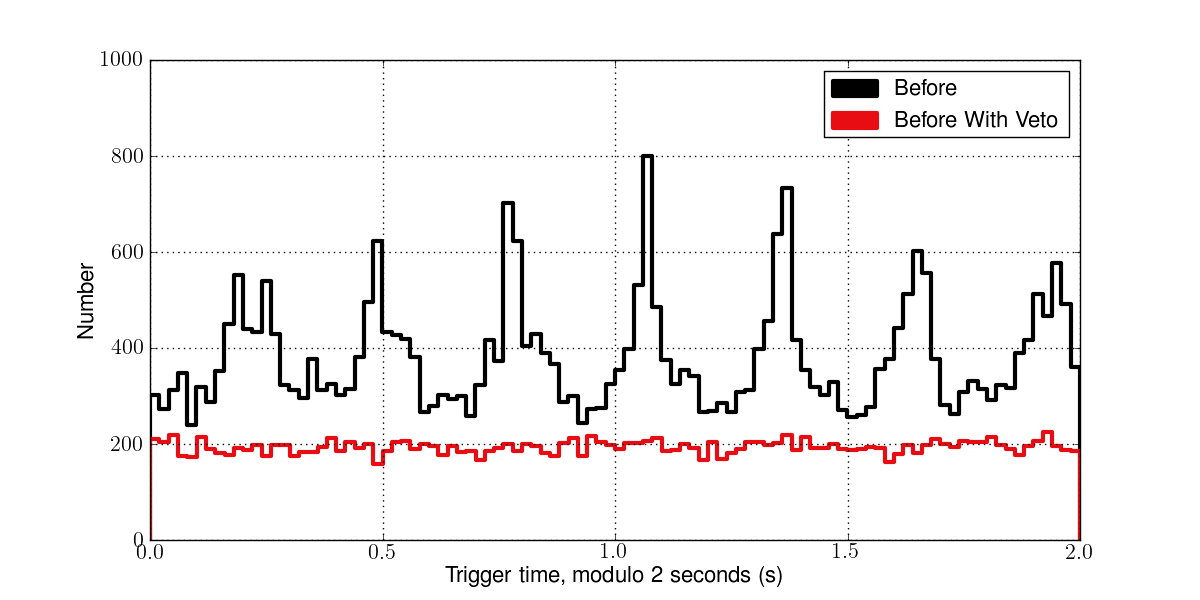}
        \label{fig:sqz_backscatter_modulo}
        }
    \caption{\label{fig:sqzbk_stacked}Triggers in a frequency band from $1\,\mathrm{kHz}$--$4\,\mathrm{kHz}$, where this noise source can be observed.
             (a) We see that in the frequency band from $1\,\mathrm{kHz}$--$4\,\mathrm{kHz}$ the effect of these low SNR glitches is more clearly visible.
             The horizontal dashed black line shows the false-alarm rate threshold of $5.6\times10^{-4}\,\mathrm{Hz}$.
             (b) Histogram of trigger time modulo 2 seconds.
             This noise source can clearly be seen in the before period, with the squeezer active, but is not observed in the before with veto period when the squeezer was turned off.}
\end{figure*}

When the $3.5\,\mathrm{Hz}$ BDO dither was active and the squeezed light injection was operating, the rate of low amplitude glitches in the shot noise limited frequency region increased, as seen in figure \ref{fig:sqz_backscatter_trig_rate-band}.
Even without commissioning work, these glitches were removed whenever the shutter between the squeezer and the interferometer was closed, and later after the $3.5\,\mathrm{Hz}$ dither was deactivated.
The glitch mechanism was not entirely clear, however there is a strong correlation of glitch occurrence with the dither period.
This can be seen in figure \ref{fig:sqz_backscatter_modulo}.
Most compellingly there was no drift in this behaviour over several months of operation, indicating that the glitch time modulation must have an origin synchronised to a digital clock, in this case, the real-time control system in which the dither sinusoid originated.
These glitches do not show up in the Omegamaps because they were of low SNR, and for the same reason they also do not affect the spectrum.

\begin{figure*}
    \centering
        \subfigure[ Strain noise spectral density]{\includegraphics[width=0.45\textwidth]{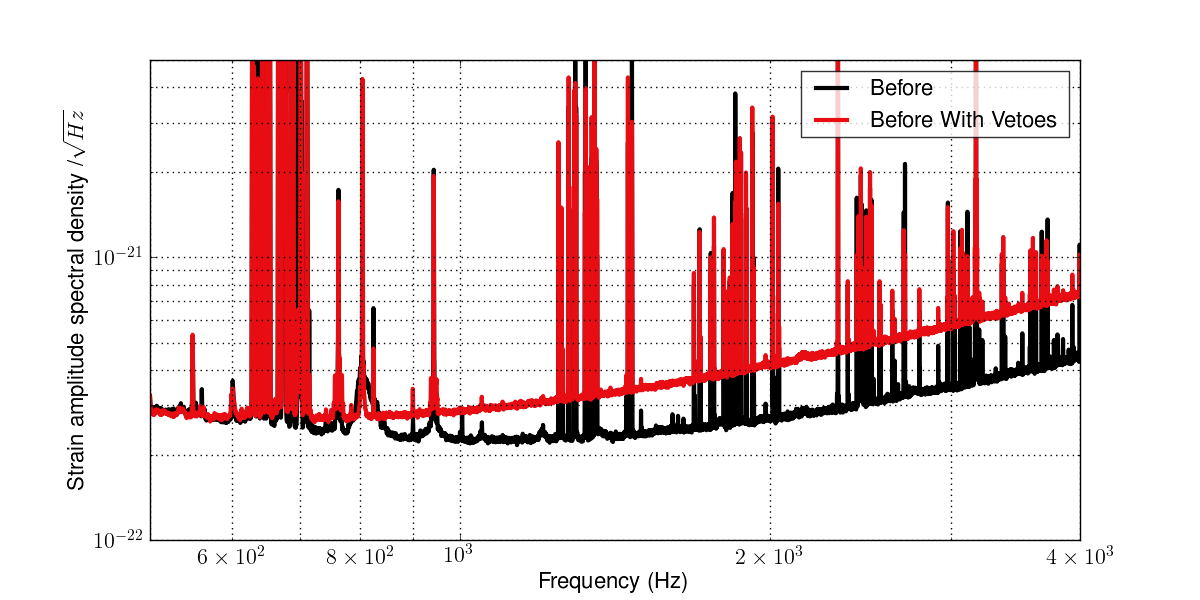}
        \label{fig:sqz_backscatter_spectrum}
        }
        \subfigure[ Cumulative glitch rate]{\includegraphics[width=0.4\textwidth]{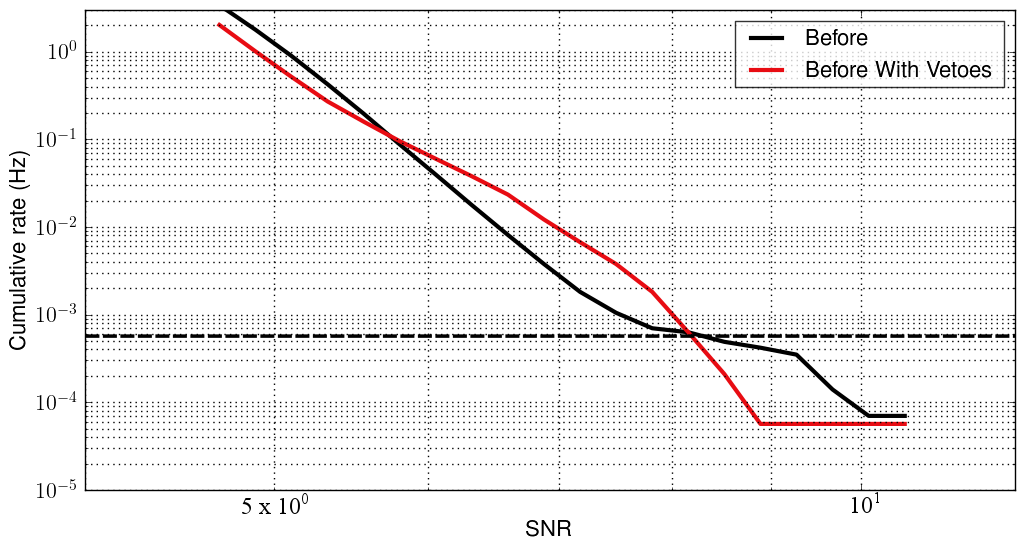}
        \label{fig:sqz_backscatter_trig_rate}
        }
    \caption{\label{fig:sqz_backscatter}$3.5\,\mathrm{Hz}$ squeezer glitches comparison plots.
             (a) A period before hypothetical commissioning with the noise source present is shown in black, and a period with out veto, removing the squeezer, is shown in red.
             (b) We see that this noise source has no notable effect on the rate of triggers in the SNR range frequency band, from $500\,\mathrm{Hz}$--$4\,\mathrm{kHz}$.
             The horizontal dashed black line shows the false-alarm rate threshold of $5.6\times10^{-4}\,\mathrm{Hz}$.}
\end{figure*}

\begin{figure}
    \centering
    \includegraphics[width=0.45\textwidth]{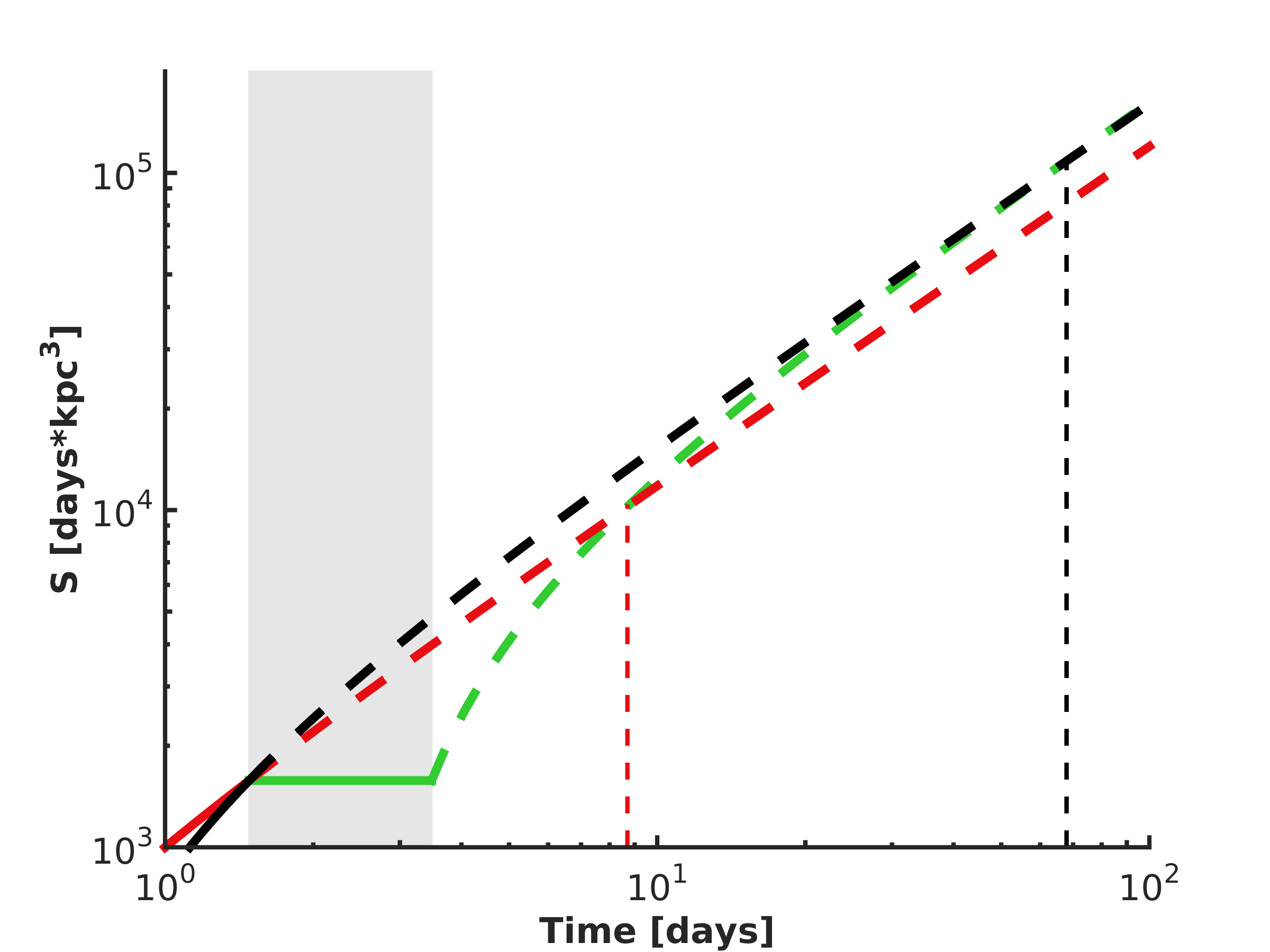}
    \caption{\label{fig:sqzbk_commvis}Graphical display of the spacetime observation volume for a GW search, $S \equiv \int \dot{S} \, \mathrm{d} t$ around the time of a hypothetical 2\,day commissioning interruption aimed at removing the $3.5\,\mathrm{Hz}$ dither squeezer glitches.
      The period displayed here starts from sometime before the commissioning period until after all benefit-delay times have been reached.
      All spacetime observation volumes are shifted so they are equal at the beginning of the commissioning period which is designated by the grey filled in area.
      Here all real spacetime observation volumes are plotted as solid lines while extrapolated or predicted spacetime observation volumes are plotted as dashed lines.
      In black we plot quantities which correspond to the before-commissioning period without vetoes while red denotes those quantities with vetoes.
      The green lines plot spacetime observation volumes during and after commissioning (hypothetical).
      The benefit-delay times are displayed by the vertical lines where the hypothetical continued before- and veto-spacetime observation rate meets the predicted-spacetime observation volume.}
\end{figure}

This noise source cannot be mitigated by the construction of a veto as the glitches occur continuously in time and broadly in frequency.
The only way to remove this glitch effect is to remove the source in the instrument or turn off the squeezer, an action which we will denote as a ``veto'' for this investigation.
Because this occurred during the S6E/VSR4 science run in 2011, detector observation time was extremely valuable and even without recourse to sophisticated astrophysical figures of merit, commissioning down-time was reserved for necessary maintenance and critical noise issues only.
At the time, only the elevated glitch rate with squeezer operation was known, and the dependence on the $3.5\,\mathrm{Hz}$ dither did not become clear until the last weeks of the science data run.
Early in the science run, the question arose as to whether to disable the squeezer, to incisively investigate the source of the glitches, or to allow squeezer operations to continue and accept the additional low SNR transients.

Fortunately during the science run, the ability to carry out studies similar to those described in this paper was available.
Elements of the coherent wave burst all-sky burst search analysis \cite{Klimenko:2008ts} were processing the coincident GEO\,600-Virgo network data on a weekly basis.
These analyses provided a figure-of-merit for the non-stationary elements of the network noise similar to $\rho$ in the integrated fixed false-alarm probability range defined by equation \ref{eq:freq_dep_range} and equation \ref{eq:integrated_range}.
A directed analysis of the $3.5\,\mathrm{Hz}$ dither squeezer glitches which compared time intervals with the squeezer shutter open and closed showed that the effect of these glitches on the network fixed false-alarm probability SNR threshold was less than a few percent.
In the single interferometer study shown in figure \ref{fig:sqz_backscatter}, we obtain similar results.
Since the squeezing at that time was acting to decrease the noise spectral density of GEO\,600 by approximately $20\%$, we decided that no intervention involving the sacrifice of science data operations was warranted.
We see the effect of this in table \ref{tab:results}; there was a $\sim 25$\% loss in the burst-like spacetime observation rate when the squeezer was off which clearly shows that deactivating the squeezer is counter productive.

To predict the performance of the detector without this noise source we combined the spectra of the detector with squeezing, and the fixed false-alarm probability SNR threshold found without squeezing.
This gives us an idea of how the detector would perform with squeezing on and without the noise source.
This gave us a predicted burst-like spacetime observation rate of $1700\,\textrm{kpc}^{3}$.

Using these values we find a benefit-delay ratio when comparing a time using squeezing to our predicted performance without the noise source of $33$; this means we would have had to find the problem and fix it within two days for us to be able to collect the benefits within the two months left in the run.
Thus performing investigations or attempting to commissioning this issue would risk inducing a deficit in potentially observed GW source which could not be recovered before the end of the S6E/VSR4 science run.
The decision to not commission was based on results from the coherent wave burst analysis performed at the time of the noise, but utilising the full benefit-delay analysis we obtain a quantitative measure of the risk that was involved in commissioning.

\begin{table*}
    \begin{center}
        \begin{tabular}{l|c|c|c|c|c}
            \hline
            \hline
            \textbf{Investigation} & \textbf{Range type} & \multicolumn{4}{c}{\textbf{Burst-like spacetime observation rates}} \\
            & \textbf{[frequency} & \textbf{Before} & \textbf{Vetoed} & \textbf{Predicted} & \textbf{After}  \\
            & \textbf{band]} & \textbf{($\times10^{5} kpc^{3}$)} & \textbf{Before} & \textbf{($\times10^{5} kpc^{3}$)} & \textbf{($\times10^{5} kpc^{3}$)} \\
            & & & \textbf{($\times10^{5} kpc^{3}$)} & & \\
            \hline
            \hline
            \multicolumn{1}{l|}{\textbf{Beam Direction}} & GRB & $6$ & $45$ & $300$ & $100$ \\
            \multicolumn{1}{l|}{\textbf{Optics (BDO)}} & [$100\,\mathrm{Hz}$--$500\,\mathrm{Hz}$] & & & & \\
            \hline
            \hline
            \multicolumn{1}{l|}{\textbf{Output mode}} & GRB & $470$ & $750$ & $850$ & $1240$ \\
            \multicolumn{1}{l|}{\textbf{cleaner (OMC)}} & [$100\,\mathrm{Hz}$--$500\,\mathrm{Hz}$] & & & & \\
            \hline
            \hline
            \multicolumn{1}{l|}{\textbf{Squeezer glitches}} & SN & $0.001$ & $0.012$ & $0.013$ & -- \\
            & [$500\,\mathrm{Hz}$--$4\,\mathrm{kHz}$] & & & & \\
            \hline
            \hline
            \multicolumn{1}{l|}{\textbf{$3.5\,\mathrm{Hz}$ dither}} & SN & $0.016$ & ${0.012}^\dagger$ & $0.017$ & -- \\
            \multicolumn{1}{l|}{\textbf{squeezer glitches}} & [$500\,\mathrm{Hz}$--$4\,\mathrm{kHz}$] & & & & \\
            \hline
            \hline
        \end{tabular}
    \end{center}
    \caption{\label{tab:results}Burst-like spacetime observation rates for each of the investigations given in section \ref{sec:investigations}.
             These spacetime observation rates are calculated from the astrophysically motivated fixed false-alarm probability ranges as described in section \ref{sec:tools} and are displayed for a period before commissioning, the same period before commissioning with vetoes applied, the predicted results of a commissioning operation, and for a period after a commissioning operation was completed.\\
             $\dagger$ The veto time for the $3.5\,\mathrm{Hz}$ dither squeezer glitches was to turn off the squeezer.}
\end{table*}

\begin{table*}
    \begin{center}
        \begin{tabular}{l|c|c|c|c|c|c}
            \hline
            \hline
            \multicolumn{1}{l|}{\textbf{Investigation}} & \textbf{Range type} &  \textbf{Commissioning} & \multicolumn{4}{c}{\textbf{Benefit-delay ratios}}\\
            & \textbf{[frequency} & \textbf{down-time} & \textbf{no vetoes,} & \textbf{no vetoes,} & \textbf{with vetoes,} & \textbf{with vetoes,}\\
            & \textbf{band]} & \textbf{(days)} & \textbf{predicted} & \textbf{actual} & \textbf{predicted} & \textbf{actual} \\
            \hline
            \hline
            \multicolumn{1}{l|}{\textbf{Beam Direction}} & GRB & 20.8 & 0.02 & 0.06 & 0.2 & 0.75\\
            \multicolumn{1}{l|}{\textbf{Optics (BDO)}} & [$100\,\mathrm{Hz}$--$500\,\mathrm{Hz}$] & & & & & \\
            \hline
            \hline
            \multicolumn{1}{l|}{\textbf{Output mode}} & GRB & 1.9 & 1.3 & 0.6 & 8 & 1.5\\
            \multicolumn{1}{l|}{\textbf{cleaner (OMC)}} & [$100\,\mathrm{Hz}$--$500\,\mathrm{Hz}$] & & & & & \\
            \hline
            \hline
            \multicolumn{1}{l|}{\textbf{Squeezer glitches}} & SN & -- & 0.08 & -- & 53 & --\\
            & [$500\,\mathrm{Hz}$--$4\,\mathrm{kHz}$] & & & & & \\
            \hline
            \hline
            \multicolumn{1}{l|}{\textbf{$3.5\,\mathrm{Hz}$ dither}} & SN & -- & 33 & -- & $2.6^\dagger$ & --\\
            \multicolumn{1}{l|}{\textbf{squeezer glitches}} & [$500\,\mathrm{Hz}$--$4\,\mathrm{kHz}$] & & & & & \\
            \hline
            \hline
        \end{tabular}
    \end{center}
    \caption{\label{tab:benefit-delay}Commissioning down-times and benefit-delay ratios for each investigation in section \ref{sec:investigations}.
             We calculate the benefit-delay ratios, as described in section \ref{sec:tools} equation \ref{eq:benefit-delay}, for scenarios where commissioning down-time is used to bring about the improvements to the spacetime observation rates shown in table \ref{tab:results}.
             Here we look at multiple cases for each investigation where we either take the spacetime observation rate calculated without vetoes or with vetoes for the era before commissioning, and either the predicted or actual spacetime observation rate for the era after commissioning.
             The total benefit-delay time can be calculated by multiplying the ratios with the actual commissioning down-times.\\
             $\dagger$ The veto time for the $3.5\,\mathrm{Hz}$ dither squeezer glitches was to turn off the squeezer.}
\end{table*}

\section{Summary}
\label{sec:summary}

As we have seen in the course of characterisation investigations of GEO\,600, different detector noise sources require different mitigation strategies to most effectively remove the noise induced reduction of the spacetime observation rate.
When these noise sources occur during a science data run, we compare the benefit-delay time when we first apply the veto solution to the remaining time in the observing period.
If the commissioning down-time plus the benefit-delay time is smaller than the remaining time in the observing period, then the commissioning should be performed to maximise the spacetime observation of the science data run.
Otherwise it is better to apply any available vetoes and then perform the commissioning after the observing period has ended.
Using this cost-benefit figure-of-merit we can quantitatively compare any available vetoes with the predicted improvement from a commissioning solution, to best inform decisions on detector commissioning.

In section \ref{sec:omc_alignment} we investigated noise in the alignment between the interferometer and the OMC.
A veto which removed the frequency bands contaminated by this noise, and a commissioning solution to remove the source of this noise were available.
In retrospect, for the BDO suspension and OMC isolation upgrades we predicted benefit-delay ratios which predicted short benefit-delay times compared to the science data run duration, and can verify that this was achieved in the end, though the actual ratios varied from the predicted by factors of a few.

In section \ref{sec:squeezer} we discuss a situation in which the glitches are short in duration and broadband, and an auxiliary channel is available from which a time segment veto can effectively be produced.
Initially the noise origin was not well known, and there was no commissioning solution available to remove this noise, but we could predict the performance of the detector without this noise source present.
The benefit-delay ratio found comparing the veto to our predicted performance without this noise source is $53$, indicating that the benefit-delay time would have been $53$ times the down-time for any commissioning.
Any investigative work for a commissioning solution that caused down-time would also require a large benefit-delay time, so it is favourable to perform non invasive investigations or work at the same time as other commissioning activities.
Therefore, applying the veto for this noise source was the best solution available.

In section \ref{sec:dither_glitches} we see that the $3.5\,\mathrm{Hz}$ dither squeezer glitches were not significantly affecting the detector's astrophysical sensitivity.
Work similar to this investigation, but instead using a network figure-of-merit which included the Virgo detector, was performed when this noise source was observed and was key in making the decision to maintain operations with the squeezer in operation, although it was the known source of this noise.
The cost-benefit analysis shows that the improvement in the spectrum, due to the squeezer, has a larger effect than the increase in the false-alarm rate threshold, due to the $3.5$\,Hz dither squeezer glitches, on the burst-like spacetime observation rate.
Therefore the best solution was continue to operate with the squeezer as it was, rather than disabling the squeezer or interrupting the observations to investigate further.

In this article, we have successfully developed, refined, and applied a cost-benefit strategy to approaching commissioning decisions aimed at solving noise sources in GEO\,600 observed during and around the time of the S6E/VSR4 data science run.
Since the development of this strategy happened only during and after the science run, the application in its final state, presented here, was only done after the fact.
For the upcoming era of advanced GW detectors, however, it is important to apply this strategy to commissioning decisions, as they arise, before and during scientific observing runs.
Accurate estimation of the commissioning time is necessary for the successful application of this strategy, which in our experience has been available from the experience of the commissioners themselves.

There is one further step which needs to be taken before application in the advanced detector network can become a reality.
All of the spacetime observation rates reported here have been calculated based on a fixed false-alarm probability range for a single detector, externally triggered, unmodelled GW burst search.
In the advanced detector era, however, the flagship GW search which is expected to make a detection is a multiple detector network search for modelled waveforms from compact binary coalescences (CBCs).
It is important that a fixed false-alarm probability range is developed for this search.
This is a non-trivial extension of the ranges used in this article, however, much progress has already been made toward reducing the computational delay of generating false-alarm probabilities for CBC triggers from a network of two or three interferometers \cite{Cannon13}.
With this hurdle overcome, it should be simple to construct a fixed false-alarm probability range for CBC searches on a network of interferometers modelled after the ranges presented in \cite{Was:2014fo}.

Once this is done, the cost-benefit strategy to commissioning decisions developed here can be carried out for commissioning decisions on individual interferometers within a network.
The effects of these decisions will be tied into the entire network by the network range figure-of-merit.
Of course with a network there will be more options for commissioning because the question may arise whether or not it makes sense to take down more than one interferometer at a time.
However, the strategy in principle remains the same.

In the advanced detector era, characterisation groups can and must evaluate the sensitivity cost of noise phenomena as they appear, develop vetoes, and compare the relative effectiveness of vetoes and commissioning solutions on the network of interferometers.
This exciting role provides a vital contribution in the preparation and operation of the first advanced detector observation runs.
With the detectors in pursuit of design sensitivity and beset by technical noises, this strategy will help to maximise the total spacetime observation volume and bring forward the first direct detection of GWs.

\section*{Acknowledgements}
We would like to thank the LIGO scientific, Virgo, and GEO collaborations for providing a medium where stimulating discussions and analyses took place which have contributed to this work.
In particular Sergey Kimenko, Gabriele Vedovato and Igor Yakushin from the coherent wave burst group ran the two detector network analysis mentioned in section \ref{sec:dither_glitches}.
The aforementioned collaborations have also allowed the use of GEO\,600 GW data published here.
The Max Planck Society, Leibniz Universitt Hannover, the Science and Technology Facilities Council in the UK, the Bundesministerium fr Bildung und Forschung and the state of Lower Saxony in Germany, and the Volkswagen Foundation all made generous contributions which helped to make GEO\,600 a reality.
Many people over many years have also worked hard on the design and construction, as well as upgrades to GEO\,600 that make it what it is today.
The GEO\,600 operators, in addition, play a large role in keeping GEO\,600 operational.
We also acknowledge the institutes out of which the authors work.
This work was supported in part by STFC Grants No. PP/F001096/1 and No. ST/I000887/1 and by the STFC Long Term Attachment ST/I505621/1.
This paper has been assigned LIGO Document No.\,P1300202.

\newpage
\bibliographystyle{unsrt}
\bibliography{references.bib}

\end{document}